\input harvmac
\input amssym

\let\includefigures=\iftrue
\let\useblackboard=\iftrue
\newfam\black

\includefigures
\message{If you do not have epsf.tex (to include figures),}
\message{change the option at the top of the tex file.}
\input epsf
\def\figin{\epsfcheck\figin}\def\figins{\epsfcheck\figins}
\def\epsfcheck{\ifx\epsfbox\UnDeFiNeD
\message{(NO epsf.tex, FIGURES WILL BE IGNORED)}
\gdef\figin##1{\vskip2in}\gdef\figins##1{\hskip.5in}
\else\message{(FIGURES WILL BE INCLUDED)}%
\gdef\figin##1{##1}\gdef\figins##1{##1}\fi}
\def\DefWarn#1{}
\def\figinsert{\goodbreak\midinsert}
\def\ifig#1#2#3{\DefWarn#1\xdef#1{fig.~\the\figno}
\writedef{#1\leftbracket fig.\noexpand~\the\figno}%
\figinsert\figin{\centerline{#3}}\medskip\centerline{\vbox{
\baselineskip12pt\advance\hsize by -1truein
\noindent\footnotefont{\bf Fig.~\the\figno:} #2}}
\endinsert\global\advance\figno by1}
\else
\def\ifig#1#2#3{\xdef#1{fig.~\the\figno}
\writedef{#1\leftbracket fig.\noexpand~\the\figno}%
\global\advance\figno by1} \fi

\def\journal#1&#2(#3){\unskip, \sl #1\ \bf #2 \rm(19#3) }
\def\andjournal#1&#2(#3){\sl #1~\bf #2 \rm (19#3) }

\def\ie{{\it i.e.}}
\def\eg{{\it e.g.}}

%
%

%


\def\unlockat{\catcode`\@=11}
\def\lockat{\catcode`\@=12}

\unlockat

\def\newsec#1{\global\advance\secno by1\message{(\the\secno. #1)}
\global\subsecno=0\global\subsubsecno=0\eqnres@t\noindent
{\bf\the\secno. #1}
\writetoca{{\secsym} {#1}}\par\nobreak\medskip\nobreak}
\global\newcount\subsecno \global\subsecno=0
\def\subsec#1{\global\advance\subsecno
by1\message{(\secsym\the\subsecno. #1)}
\ifnum\lastpenalty>9000\else\bigbreak\fi\global\subsubsecno=0
\noindent{\it\secsym\the\subsecno. #1}
\writetoca{\string\quad {\secsym\the\subsecno.} {#1}}
\par\nobreak\medskip\nobreak}
\global\newcount\subsubsecno \global\subsubsecno=0
\def\subsubsec#1{\global\advance\subsubsecno by1
\message{(\secsym\the\subsecno.\the\subsubsecno. #1)}
\ifnum\lastpenalty>9000\else\bigbreak\fi
\noindent\quad{\secsym\the\subsecno.\the\subsubsecno.}{#1}
\writetoca{\string\qquad{\secsym\the\subsecno.\the\subsubsecno.}{#1}}
\par\nobreak\medskip\nobreak}

\def\subsubseclab#1{\DefWarn#1\xdef
#1{\noexpand\hyperref{}{subsubsection}%
{\secsym\the\subsecno.\the\subsubsecno}%
{\secsym\the\subsecno.\the\subsubsecno}}%
\writedef{#1\leftbracket#1}\wrlabeL{#1=#1}}
\lockat

\def\ie{{\it i.e.}}
\def\eg{{\it e.g.}}

\def\CM {{\cal M}}
\def\CN {{\cal N}}

\font\manual=manfnt \def\dbend{\lower3.5pt\hbox{\manual\char127}}

\def\IZ{\relax\ifmmode\mathchoice
{\hbox{\cmss Z\kern-.4em Z}}{\hbox{\cmss Z\kern-.4em Z}}
{\lower.9pt\hbox{\cmsss Z\kern-.4em Z}}
{\lower1.2pt\hbox{\cmsss Z\kern-.4em Z}}\else{\cmss Z\kern-.4em
Z}\fi}
\def\half{{1\over 2}}

\def\CM {{\cal M}}
\def\CN {{\cal N}}


\def\IZ{\relax\ifmmode\mathchoice
{\hbox{\cmss Z\kern-.4em Z}}{\hbox{\cmss Z\kern-.4em Z}}
{\lower.9pt\hbox{\cmsss Z\kern-.4em Z}}
{\lower1.2pt\hbox{\cmsss Z\kern-.4em Z}}\else{\cmss Z\kern-.4em
Z}\fi}
\def\IB{\relax{\rm I\kern-.18em B}}
\def\IC{{\relax\hbox{$\inbar\kern-.3em{\rm C}$}}}
\def\ID{\relax{\rm I\kern-.18em D}}
\def\IE{\relax{\rm I\kern-.18em E}}
\def\IF{\relax{\rm I\kern-.18em F}}
\def\IG{\relax\hbox{$\inbar\kern-.3em{\rm G}$}}
\def\IGa{\relax\hbox{${\rm I}\kern-.18em\Gamma$}}
\def\IH{\relax{\rm I\kern-.18em H}}
\def\II{\relax{\rm I\kern-.18em I}}
\def\IK{\relax{\rm I\kern-.18em K}}
\def\IP{\relax{\rm I\kern-.18em P}}
\def\IQ{\relax\hbox{$\inbar\kern-.3em{\rm Q}$}}

\def\inbar{\,\vrule height1.5ex width.4pt depth0pt}

\font\cmss=cmss10 \font\cmsss=cmss10 at 7pt
\def\IR{\relax{\rm I\kern-.18em R}}

\def\Tr{\rm Tr}

%
%

\def\makeblankbox#1#2{\hbox{\lower\dp0\vbox{\hidehrule{#1}{#2}%
   \kern -#1
   \hbox to \wd0{\hidevrule{#1}{#2}%
      \raise\ht0\vbox to #1{}
      \lower\dp0\vtop to #1{}
      \hfil\hidevrule{#2}{#1}}%
   \kern-#1\hidehrule{#2}{#1}}}%
}%
\def\hidehrule#1#2{\kern-#1\hrule height#1 depth#2 \kern-#2}%
\def\hidevrule#1#2{\kern-#1{\dimen0=#1\advance\dimen0 by #2\vrule
    width\dimen0}\kern-#2}%
\def\openbox{\ht0=1.2mm \dp0=1.2mm \wd0=2.4mm  \raise 2.75pt
\makeblankbox {.25pt} {.25pt}  }

\def\bun#1/#2{\leavevmode
   \kern.1em \raise .5ex \hbox{\the\scriptfont0 #1}%
   \kern-.1em $/$%
   \kern-.15em \lower .25ex \hbox{\the\scriptfont0 #2}%
}

\def\opensquare{\ht0=3.4mm \dp0=3.4mm \wd0=6.8mm  \raise 2.7pt
\makeblankbox {.25pt} {.25pt}  }


\def\sector#1#2{\ {\scriptstyle #1}\hskip 1mm
\mathop{\opensquare}\limits_{\lower 1mm\hbox{$\scriptstyle#2$}}\hskip 1mm}

\def\tsector#1#2{\ {\scriptstyle #1}\hskip 1mm
\mathop{\opensquare}\limits_{\lower 1mm\hbox{$\scriptstyle#2$}}^\sim\hskip 1mm}


\def\inbar{\,\vrule height1.5ex width.4pt depth0pt}

\font\cmss=cmss10 \font\cmsss=cmss10 at 7pt
\def\IR{\relax{\rm I\kern-.18em R}}

\def\Tr{\rm Tr}


\def\frac#1#2{{#1\over#2}}

\def\half{\frac12}

\def\inbar{\,\vrule height1.5ex width.4pt depth0pt}
\def\IC{\relax\hbox{$\inbar\kern-.3em{\rm C}$}}
\def\IR{\relax{\rm I\kern-.18em R}}
\def\IP{\relax{\rm I\kern-.18em P}}

%
%
\catcode`\@=11
\def\slash#1{\mathord{\mathpalette\c@ncel{#1}}}
\overfullrule=0pt

\def\II{{\cal I}}

\def\underrel#1\over#2{\mathrel{\mathop{\kern\z@#1}\limits_{#2}}}

\catcode`\@=12


%

\def\det{{\rm det}}

\def\cosh{{\rm cosh}}

\def\det{{\rm det}}
\def\exp{{\rm exp}}



\def\frac#1#2{{#1\over#2}}

\def\half{\frac12}

\def\inbar{\,\vrule height1.5ex width.4pt depth0pt}
\def\IC{\relax\hbox{$\inbar\kern-.3em{\rm C}$}}
\def\IR{\relax{\rm I\kern-.18em R}}
\def\IP{\relax{\rm I\kern-.18em P}}

%
%

%
\catcode`\@=11
\def\slash#1{\mathord{\mathpalette\c@ncel{#1}}}
\overfullrule=0pt

\def\II{{\cal I}}

\def\underrel#1\over#2{\mathrel{\mathop{\kern\z@#1}\limits_{#2}}}

\catcode`\@=12
\def\IZ{\Bbb Z}

%

\def\det{{\rm det}}

\def \cosh{{\rm cosh}}

\def\det{{\rm det}}
\def\exp{{\rm exp}}


\def\id{\protect{{1 \kern-.28em {\rm l}}}}
\def\al{\alpha}

\lref\SeibergBX{N.~Seiberg,
``Observations on the moduli space of two dimensional string theory,''
JHEP {\bf 0503}, 010 (2005)
[arXiv:hep-th/0502156].
}

\lref\SuyamaXK{
  T.~Suyama,
  ``Deformation of CHS model,''
  Nucl.\ Phys.\ B {\bf 641}, 341 (2002)
  [arXiv:hep-th/0206171].
}

\lref\BergmanQF{
  O.~Bergman and S.~Hirano,
  ``Semi-localized instability of the Kaluza-Klein linear dilaton vacuum,''
  arXiv:hep-th/0510076.
}

\lref\CallanAT{ C.~G.~Callan, J.~A.~Harvey and A.~Strominger,
``Supersymmetric string solitons,'' arXiv:hep-th/9112030.
}

\lref\AharonyUB{ O.~Aharony, M.~Berkooz, D.~Kutasov and
N.~Seiberg, ``Linear dilatons, NS5-branes and holography,'' JHEP
{\bf 9810}, 004 (1998) [arXiv:hep-th/9808149].
}

\lref\KutasovUA{ D.~Kutasov and N.~Seiberg, ``Noncritical
Superstrings,'' Phys.\ Lett.\ B {\bf 251}, 67 (1990).
}

\lref\GiveonFU{ A.~Giveon, M.~Porrati and E.~Rabinovici, ``Target
space duality in string theory,'' Phys.\ Rept.\  {\bf 244}, 77
(1994) [arXiv:hep-th/9401139].
}

\lref\ItzhakiDD{ N.~Itzhaki, J.~M.~Maldacena, J.~Sonnenschein and
S.~Yankielowicz, ``Supergravity and the large N limit of theories
with sixteen  supercharges,'' Phys.\ Rev.\ D {\bf 58}, 046004
(1998) [arXiv:hep-th/9802042].
}

\lref\ItzhakiTU{ N.~Itzhaki, N.~Seiberg and D.~Kutasov, ``I-brane
dynamics,'' arXiv:hep-th/0508025.
}

\lref\GiveonZM{ A.~Giveon, D.~Kutasov and O.~Pelc, ``Holography
for non-critical superstrings,'' JHEP {\bf 9910}, 035 (1999)
[arXiv:hep-th/9907178].
}

\lref\PolchinskiRR{ J.~Polchinski, ``String theory. Vol. 2:
Superstring theory and beyond''  Cambridge press.
 }

\lref\pol{J. Polchinski, "String Theory, Vol II" Cambridge press.
}

\lref\LinNH{ H.~Lin and J.~Maldacena, ``Fivebranes from gauge
theory,'' arXiv:hep-th/0509235.
}

 \lref\KutasovSV{ D.~Kutasov and N.~Seiberg,
``Number Of Degrees Of Freedom, Density Of States And Tachyons In
String Theory And Cft,'' Nucl.\ Phys.\ B {\bf 358}, 600 (1991).
}

\lref\KiritsisIU{
  E.~Kiritsis and C.~Kounnas,
  ``Infrared behavior of closed superstrings in strong magnetic and
  Nucl.\ Phys.\ B {\bf 456}, 699 (1995)
  [arXiv:hep-th/9508078].
}

\lref\ForsteKM{
  S.~Forste and D.~Roggenkamp,
  ``Current current deformations of conformal field theories, and WZW
  JHEP {\bf 0305}, 071 (2003)
  [arXiv:hep-th/0304234].
}

\lref\HassanMQ{
  S.~F.~Hassan and A.~Sen,
  ``Twisting classical solutions in heterotic string theory,''
  Nucl.\ Phys.\ B {\bf 375}, 103 (1992)
  [arXiv:hep-th/9109038].
}

\lref\AharonyVK{ O.~Aharony, B.~Fiol, D.~Kutasov and
D.~A.~Sahakyan, ``Little string theory and heterotic/type II
duality,'' Nucl.\ Phys.\ B {\bf 679}, 3 (2004)
[arXiv:hep-th/0310197].
}

\lref\AdamsSV{ A.~Adams, J.~Polchinski and E.~Silverstein, ``Don't
panic! Closed string tachyons in ALE space-times,'' JHEP {\bf
0110}, 029 (2001) [arXiv:hep-th/0108075].
}

\lref\HassanGI{ S.~F.~Hassan and A.~Sen, ``Marginal deformations
of WZNW and coset models from O(d,d) transformation,'' Nucl.\
Phys.\ B {\bf 405}, 143 (1993) [arXiv:hep-th/9210121].
}

\lref\GiveonPH{ A.~Giveon and E.~Kiritsis, ``Axial vector duality
as a gauge symmetry and topology change in string theory,'' Nucl.\
Phys.\ B {\bf 411}, 487 (1994) [arXiv:hep-th/9303016].
}

\lref\BanksYZ{
T.~Banks and L.~J.~Dixon,
``Constraints On String Vacua With Space-Time Supersymmetry,''
Nucl.\ Phys.\ B {\bf 307}, 93 (1988).
}

\lref\SenNU{
  A.~Sen,
  ``Rolling tachyon,''
  JHEP {\bf 0204}, 048 (2002)
  [arXiv:hep-th/0203211].
}

\lref\GutperleAI{ M.~Gutperle and A.~Strominger, ``Spacelike
branes,'' JHEP {\bf 0204}, 018 (2002) [arXiv:hep-th/0202210].
}

\lref\LambertZR{
  N.~Lambert, H.~Liu and J.~Maldacena,
  ``Closed strings from decaying D-branes,''
  arXiv:hep-th/0303139.
}

\lref\GaiottoRM{ D.~Gaiotto, N.~Itzhaki and L.~Rastelli, ``Closed
strings as imaginary D-branes,'' Nucl.\ Phys.\ B {\bf 688}, 70
(2004) [arXiv:hep-th/0304192].
}

\lref\GiveonMI{
A.~Giveon, D.~Kutasov, E.~Rabinovici and A.~Sever,
``Phases of quantum gravity in AdS(3) and linear dilaton backgrounds,''
Nucl.\ Phys.\ B {\bf 719}, 3 (2005)
[arXiv:hep-th/0503121].
}

\lref\DineVU{
M.~Dine, P.~Y.~Huet and N.~Seiberg,
``Large And Small Radius In String Theory,''
Nucl.\ Phys.\ B {\bf 322}, 301 (1989).
}

\lref\GiveonPX{
  A.~Giveon and D.~Kutasov,
  ``Little string theory in a double scaling limit,''
  JHEP {\bf 9910}, 034 (1999)
  [arXiv:hep-th/9909110].
}

\lref\GiveonTQ{
  A.~Giveon and D.~Kutasov,
  ``Comments on double scaled little string theory,''
  JHEP {\bf 0001}, 023 (2000)
  [arXiv:hep-th/9911039].
}


\lref\DineVF{ M.~Dine and N.~Seiberg, ``Microscopic Knowledge From
Macroscopic Physics In String Theory,'' Nucl.\ Phys.\ B {\bf 301},
357 (1988).
}

\lref\HarveyWM{
  J.~A.~Harvey, D.~Kutasov, E.~J.~Martinec and G.~W.~Moore,
  ``Localized tachyons and RG flows,''
  arXiv:hep-th/0111154.
}

\lref\MurthyES{ S.~Murthy, ``Notes on non-critical superstrings in
various dimensions,'' JHEP {\bf 0311}, 056 (2003)
[arXiv:hep-th/0305197].
}

\lref\KarczmarekBW{
  J.~L.~Karczmarek, J.~Maldacena and A.~Strominger,
  ``Black hole non-formation in the matrix model,''
  arXiv:hep-th/0411174.
}

\Title{\vbox{\baselineskip12pt\hbox{hep-th/0510087}
\hbox{PUPT/2177}}} {\vbox{
\centerline{Non-Supersymmetric Deformations of}
\bigskip
\centerline{ Non-Critical Superstrings} }}
\medskip

\centerline{\it Nissan Itzhaki $ ^{a,b}$, David Kutasov $ ^c$ and
Nathan Seiberg $ ^b$}
\bigskip
\centerline{$ ^a$Department of Physics, Princeton University,
Princeton, NJ 08544}
\medskip
\centerline{$ ^b$School of Natural Sciences, Institute for
Advanced Study} \centerline{Einstein Drive, Princeton, NJ 08540}
\medskip
\centerline{$ ^c$EFI and Department of Physics, University of
Chicago}\centerline{5640 S. Ellis Av. Chicago, IL 60637}

\smallskip

\vglue .3cm

\bigskip

\bigskip
 \noindent
We study certain supersymmetry breaking deformations of linear
dilaton backgrounds in different dimensions. In some cases, the
deformed theory has bulk closed strings tachyons. In other cases
there are no bulk tachyons, but there are localized tachyons. The
real time condensation of these localized tachyons is described by
an exactly solvable worldsheet CFT. We also find some stable,
non-supersymmetric backgrounds.

\bigskip

\vglue .3cm
\bigskip

\Date{10/05}

\newsec{Introduction and summary}

In this paper we will study some aspects of string propagation in
asymptotically linear dilaton spacetimes. These backgrounds have a
boundary at infinity in a spatial direction, $\phi$, near which
they look like \eqn\aslindil{\IR^{d}\times\IR_\phi\times \CN~.}
The signature of $\IR^d$ is either  Lorentzian  or Euclidean.
$\IR_\phi$ is the real line labelled by $\phi$, and $\CN$ is a
compact space. The dilaton varies with $\phi$ as follows:
\eqn\dilslope{\Phi=-{Q\over2}\phi~,} where we set $\alpha'=2$, so
the worldsheet central charge of $\phi$ is $c_\phi=1+3Q^2$. The
string coupling $g_s=e^{\Phi}$ vanishes at the boundary
$\phi=\infty$ and grows as we move away from it.

Much of the work on backgrounds of the form \aslindil\ in recent
years focused on solutions that preserve spacetime supersymmetry.
The main purpose of this paper is to study non-supersymmetric
backgrounds of the form \aslindil. We will consider, following
\KutasovUA, solutions of the form
\eqn\ksnon{\IR^{d}\times\IR_\phi\times S^1\times\CM/\Gamma~,}
where $\CM$ is a CFT with $N=2$ worldsheet supersymmetry, and
$\Gamma$ is a discrete group associated with the chiral GSO
projection, which acts on $S^1\times\CM$.  When the radius of
$S^1$ is equal to $Q$, the background \ksnon\ is spacetime
supersymmetric. It was noted in \KutasovUA\ that changing the
radius of the $S^1$ provides a natural way of breaking spacetime
supersymmetry continuously (while evading the no-go theorem of
\refs{\BanksYZ,\DineVF}). We will study the resulting moduli space
of vacua. For $d=0$ and no $\CM $ (the empty theory), this was
done in \SeibergBX. The main new issue that needs to be addressed
for $d>0$ (or for $d=0$ and non-trivial $\CM$) is the stability of
the solution along the moduli space. This will be the focus of our
discussion.

There are in fact two kinds of instabilities that can appear in
spacetimes of the form \aslindil, corresponding to the two kinds
of physical states  in such spacetimes. One kind is delta function
normalizable states  in the bulk of the linear dilaton throat
$\IR_\phi$. These states are characterized by their $\phi$
momentum, $p$. Their wavefunctions behave like $\exp(ip\phi)$. The
other is normalizable states localized deep inside the throat.
Their wavefunctions decay at large $\phi$ like $\exp(-m\phi)$.

The two kinds of states have a natural interpretation from the
spacetime point of view. Backgrounds of the form \aslindil,
\ksnon\ typically appear in the vicinity of fivebranes or
singularities in string theory
\refs{\CallanAT\ItzhakiDD\AharonyUB\GiveonZM-\MurthyES}. In these
geometries, the delta function normalizable states correspond to
bulk modes propagating away from the singularity, while the
normalizable ones correspond to states localized at the
singularity. They are roughly analogous to untwisted and twisted
states in a non-compact orbifold. When the theory is
supersymmetric these two kinds of states are non-tachyonic.  But
when we break supersymmetry  one or both of them can become
tachyonic. We will analyze backgrounds of the form \aslindil\ in
different dimensions, with a particular focus on the question
whether they exhibit instabilities of either kind when
supersymmetry is broken.

We will see that the stability properties depend on the dimension $d$
and compact manifold $\CN$. After discussing some general features
in section 2, we turn to examples.
In section 3 we consider the case $d=4$  with $\CM=0$ in \ksnon,
which corresponds to the near-horizon geometry of the conifold, or
equivalently two $NS5$-branes intersecting on an $\IR^{3,1}$. We
describe the moduli space of non-supersymmetric vacua, and find
that everywhere except at the supersymmetric point there is a tachyon
in the bulk of the linear dilaton throat.

In sections 4 and 5 we study the case $d=6$, with $\CN=SU(2)_k$.
{}From a ten dimensional perspective, this is the near-horizon
geometry of $k$ parallel $NS5$-branes. The linear dilaton slope
\dilslope\ is given by $Q=\sqrt{2/k}$. In this case changing $R$
corresponds to adding an exactly marginal deformation that breaks
the $SU(2)_L\times SU(2)_R$ symmetry associated with the CHS
background \CallanAT. We find that if $R$ is not too far from $Q$,
there are no tachyonic bulk modes. The reason for that is that at
the supersymmetric point, the bulk modes form a continuum above a
gap. Thus, to get tachyons in the bulk one needs to move a finite
distance away from the supersymmetric point.

Despite the absence of bulk tachyons, these six dimensional
non-supersymmetric backgrounds are unstable, since for all $R\neq Q$
the system has localized tachyons. These have a natural fivebrane
interpretation. At the supersymmetric point, there are four flat
directions in field space associated with the positions of the
fivebranes in the transverse space. These directions are flat
since the gravitational attraction of the fivebranes is precisely
cancelled by their repulsion due to the  NS $B$-field. Away from
the supersymmetric point, the cancellation between the
gravitational attraction and the $B$-field repulsion is spoiled,
and the position of the fivebranes in the different directions
develops a potential. Two of the directions become massive (\ie\
fivebranes separated in these directions attract) while the other
two become tachyonic (corresponding to repulsive interactions
between the fivebranes). We find the exact worldsheet CFT that
corresponds to fivebranes rotating on a circle in the directions
in which they attract, and the exact CFT  which describes  them
running away to infinity in the directions in which they repel. In
the appendix, we construct the  supergravity solutions associated
with these CFT's.

In section 6 we consider the case $d=2$, $\CM=0$ \ksnon. As in the
$d=6$ case, there is a finite interval around the supersymmetric
point where no bulk tachyons appear. However, in this case there
are no localized tachyons either. Thus, this background is an
example of a stable, non-supersymmetric deformation of
non-critical superstrings. In fact, it is a special case of a more
general phenomenon. It was pointed out in \refs{\KarczmarekBW,
\GiveonMI} that non-critical superstring backgrounds with $Q^2>2$
are in a different phase than those with $Q^2<2$. While the latter
have the property that the high energy spectrum is dominated by
non-perturbative states (two dimensional black holes), for the
former the high energy spectrum is perturbative \GiveonMI. Here we
find that there is a difference in the infrared behavior as well.
While for $Q^2<2$, the lowest lying states for $R\not=Q$ are
tachyons that are localized deep in the strong coupling region of
the throat, for $Q^2>2$ such tachyons are absent. This suggests a
kind of UV/IR connection beyond the one familiar in perturbative
string theory \KutasovSV.

The stable, non-supersymmetric models discussed in section 6 all
have a linear dilaton slope of order one in string units. In section 7
we consider a three dimensional background, which has the form
\aslindil, is stable after supersymmetry breaking, but has the property
that the slope of the dilaton can be arbitrarily small. This is possible
since, as discussed in \refs{\ItzhakiTU,\LinNH} at the supersymmetric
point this model has a finite mass gap.

Other aspects of instabilities in linear dilaton backgrounds were
recently considered in \BergmanQF.

\bigskip
\vbox{
$$\vbox{\offinterlineskip
 \hrule height 1.1pt \halign{&\vrule width 1.1pt#
 &\strut\quad#\hfil\quad& \vrule width 1.1pt#
 &\strut\quad#\hfil\quad& \vrule width 1.1pt#
 &\strut\quad#\hfil\quad& \vrule width 1.1pt#
 &\strut\quad#\hfil\quad& \vrule width 1.1pt#\cr
 height0pt &\omit&&\omit& &\omit& \cr &\hfil & &\hfil bulk stability &
 &\hfil localized stability & &\hfill arbitrarily small  slope & \cr
 height0pt &\omit& &\omit& &\omit& \cr
 \noalign{\hrule height 1.1pt} height0pt &\omit& &\omit& &\omit& \cr
 &\hfil $d=4$& &\hfil $-$& &\hfil $-$ & &\hfil $-$ &\cr
 height0pt &\omit& &\omit& &\omit& \cr
 \noalign{\hrule} height0pt &\omit& &\omit& &\omit& \cr
 &\hfil $d=6$& &\hfil $+$& &\hfil $-$& &\hfil $-$ &\cr
 height0pt &\omit& &\omit& &\omit& \cr
 \noalign{\hrule} height0pt &\omit& &\omit& &\omit& \cr
 &\hfil $d=2$& &\hfil $+$& &\hfil $+$& &\hfil $-$& \cr
 height0pt &\omit& &\omit& &\omit& \cr
 \noalign{\hrule} height0pt &\omit& &\omit& &\omit& \cr
 &\hfil $d=3$& &\hfil $+$& &\hfil $+$& &\hfil $+$& \cr
 height0pt &\omit& &\omit& &\omit& \cr
  }
 \hrule height 1.1pt }$$
} \centerline{\rm Table 1: Stability properties of the different
backgrounds considered in the paper.}

\newsec{ Generalities }

The $\IR_\phi\times S^1$ part of \ksnon\ is described by two $(1,1)$
worldsheet superfields, $\Phi$ and $X$, whose component form is
 \eqn\super{\eqalign{X&=x+\theta \psi_x +\bar{\theta}
 \bar{\psi}_x +\theta \bar{\theta} F_x ~,\cr
 \Phi &= \phi +\theta \psi +\bar{\theta}
 \bar{\psi} +\theta \bar{\theta} F~. \cr}}
The linear dilaton in the $\phi$ direction, \dilslope, implies
that the central charge of the two superfields \super\ is
$\hat{c}_L=2+2 Q^2$. To form a background of string theory, we add
to \super\ the worldsheet theory corresponding to $\IR^{d-1,1}$
and the compact CFT $\CM$  such that the total $\hat{c} =10$. The
starting point of our discussion is the spacetime supersymmetric
theory of \KutasovUA. That theory contains the spacetime
supercharges \eqn\spacetimesup{\eqalign{ Q_\alpha^+=&\oint{dz\over
2\pi i} e^{-{\varphi\over2}}e^{-{i\over2}(H+aZ-Qx)}S_\alpha~,\cr
Q_{\bar\alpha}^-=&\oint{dz\over 2\pi i}
e^{-{\varphi\over2}}e^{{i\over2}(H+aZ-Qx)}S_{\bar\alpha}~.\cr }}
The notation we use  is the following (see
\refs{\KutasovUA,\GiveonZM} for more detail). $\varphi$ is the
bosonized superconformal ghost. $S_\alpha$ and $S_{\bar\alpha}$
are spin fields of ${\rm Spin}(d-1,1)$. For $d=2,6$ they transform
in the same spinor representation, while for $d=4,8$ they are in
different representations. $H$ is a compact scalar field which
bosonizes $\psi_x$, $\psi$ \super. The constant $a$ is related to
the central charge of the compact CFT $\CM$ in \ksnon,
$a=\sqrt{c_\CM/3}$. The field $Z$ bosonizes the $U(1)_R$ current
in the $N=2$ superconformal algebra of $\CM$. A similar set of
spacetime supercharges arises from the other worldsheet chirality.

The fact that the supercharges \spacetimesup\ carry momentum and
winding in the $x$ direction means that we should think of $x$ as
a compact field. In order for the GSO projection to act as a
$Z_2$, we must take the radius of $x$ to be $Q$ or $2/Q$ (the two
are related by T-duality). A natural question is whether
increasing and decreasing the radius from the supersymmetric point
are equivalent operations. The two are related by changing the
sign of  the left-moving part of $x$, and $\psi_x$,  while leaving
the right-movers intact. In order to see how this transformation
acts on the non-critical superstring, we need to examine its
action on the spacetime supercharges (as in \DineVU). In the
notation of equation \spacetimesup, this transformation takes
$(H,x)\to-(H,x)$, and acts on the compact CFT $\CM$ as well,
taking $Z\to-Z$. If $S_\alpha$  and $S_{\bar\alpha}$  belong to
the same spinor representation of ${\rm Spin}(d-1,1)$ (which is
the case for $d=2,6$), this transformation simply exchanges the
two lines of \spacetimesup, and we conclude that the
supersymmetric theory is self-dual, so that increasing and
decreasing $R$ gives rise to isomorphic theories. Otherwise (for
$d=4,8$), this transformation maps IIA to IIB, which are not
isomorphic, like in the critical string \DineVU\ (which
corresponds to $d=8$ in our notations).

A special case of the general construction corresponds to no $\CM$
in \ksnon. The resulting spacetime is \eqn\irdd{\IR^{d-1,1}\times
S^1 \times \IR_{\phi}~,} and $Q=\sqrt{4-{d\over2}}$. The geometry
\irdd\ arises in the vicinity of a singularity of the form
\eqn\singcy{z_1^2+z_2^2+\cdots +z_{n+1}^2=0} of a Calabi-Yau
$n$-fold, with $2n+d=10$ \GiveonZM. In particular, the $d=0$ case
discussed in \SeibergBX\ corresponds to a singularity of a
Calabi-Yau fivefold. It is believed that \irdd, \singcy\ can be
alternatively interpreted as the near-horizon geometry of an
$NS5$-brane wrapped around the surface
\eqn\wrappedfive{z_1^2+z_2^2+\cdots+z_{n-1}^2=0~.} For $d=6$,
\wrappedfive\ describes two fivebranes located at the point
$z_1=0$. For $d=4$, it describes two fivebranes intersecting along
$\IR^{3,1}$. For $d=2,0$ one has a fivebrane wrapping an $A_1$ ALE
space and a conifold, respectively. The latter can be described in
terms of intersecting fivebranes as well, by applying further
T-duality in an angular direction of the cone \wrappedfive .

The lowest lying bulk states that propagate in  \irdd\ are closed
string tachyons, carrying either momentum or winding in the $x$
direction. Their vertex operators have the form
\eqn\vertach{e^{-\varphi+ip_Lx_L+ip_Rx_R+ip_\mu
x^\mu+\beta\phi}~.} The mass-shell condition is
\eqn\masss{M^2={Q^2\over4}-1+p_L^2+\lambda^2=
{Q^2\over4}-1+p_R^2+\lambda^2~,} where $\lambda$ is the momentum
in the $\phi$ direction, $\beta=-{Q\over2}+i\lambda$.

At the supersymmetric point, the GSO projection requires that the
left and right moving momenta in the $x$ direction satisfy the
constraint $Qp_{L,R}\in 2\IZ +1$. For the momentum modes,
$p_L=p_R={n\over Q}$, and the above constraint implies that $n\in
2\IZ +1$. As we change the radius, one has $p_L=p_R={n\over R}$
and \masss\ implies that the lowest momentum tachyon (which has
$n=1$) is non-tachyonic when \eqn\nontachmom{{1\over R^2}\ge
1-{Q^2\over4}={d\over8}~.} For the winding modes, at the
supersymmetric point one has $p_L=-p_R={wQ\over2}$, and the GSO
constraint implies that \eqn\windgso{\half w Q^2\in 2\IZ+1~,} or
\eqn\rangew{w\in {4\IZ+2\over 4-{d\over2}}~.} Note that for $d=2$
the winding $w$ is in general fractional\foot{It is possible to
rescale $R$ such that $w\in \IZ$.}. In the T-dual picture, where
the supersymmetric radius is $2/Q$ and the roles of momentum and
winding are exchanged, the winding is always integer and the
momentum can be fractional.

The condition that the tachyon with the lowest winding,
$w=2/(4-{d\over2})$, be non-tachyonic takes the form
\eqn\nontachwind{R^2\ge {d\over8}Q^4~.}
Combining the conditions \nontachmom, \nontachwind\ we conclude that
for $d=4$ the theory contains a bulk tachyon for any $R\not=Q$, whereas
for $d=2, 6$ there is a finite range in which the bulk spectrum is
non-tachyonic:
\eqn\rangets{\eqalign{
d=2:& \qquad {3\over2}\le R\le 2~,\cr
d=6:& \qquad {\sqrt3\over2}\le R\le {2\over\sqrt{3}}~.\cr
}}
In general, the spectrum of the theory in the range \rangets\ exhibits
a finite gap, which goes to zero as we approach the endpoints.

The line of theories labelled by the radius $R$ which is discussed
above is expected, as in the $d=0$ case of \SeibergBX, to be part
of a richer moduli space of theories which can be obtained by
modding out by discrete symmetries. In the next sections we will
study the resulting moduli spaces for the different cases, and in
particular their stability properties.

An interesting property of these moduli spaces is that when the
radius $R\to\infty$, all spacetime fermions are lifted to infinite
mass, and one finds the non-critical $0A$ and $0B$ theories. This
is different from the situation in the critical and two
dimensional string (which correspond in our notations to $d=8$ and
$d=0$, respectively). The reason  is that in the non-compact case,
spacetime fermions do not satisfy level matching. Consider, for
example, the $(NS,+;R,\pm)$ sector. All states in this sector
satisfy \eqn\nsrplus{L_0-\bar L_0\in {1\over 2}-{d\over16}+\IZ~.}
This is consistent with level matching for $d=8$ (the critical
string) but not for lower dimensions. Similarly, in the
$(NS,-;R,\pm)$ sector, one has \eqn\nsrminus{L_0-\bar L_0\in
-{d\over16}+\IZ~,} and one can level match in $d=0$ (the two
dimensional string), but not in higher dimensions. Therefore, for
$d=2,4,6$ there are no spacetime fermions for infinite radius $R$,
and the theory reduces to type 0.

The bulk modes are described by a free worldsheet theory, and thus
can be analyzed using the same methods as in the critical
superstring. We will follow the notation of \PolchinskiRR\   and
denote the closed string sectors by
 \eqn\eqss{ (\al, F, \bar{\al}, \bar{F} )~.}
$\al$ is the space-time fermion number; it is $0$ in the NS sector
and $1$ in the R sector. $F$ is the world sheet fermion number. In
the NS sector, it is $0$ in the $+$ sector (\eg\ for the graviton)
and $1$ in the $-$ sector (\eg\ for the tachyon). In the Ramond
sector, its possible values depend on the dimension. For $d=4k$,
it takes the values $0$ and $1$, whereas for $d=2+4k$, it takes
the values $\pm\half$. Both $\alpha$ and $F$ are defined modulo
two.

\newsec{Four dimensional background}

In this section we study string propagation in $\IR^{3,1} \times
\IR_{\phi}\times S^1$.  The dilaton slope \dilslope\  in this case
is $Q=\sqrt2$. From a ten dimensional perspective this theory
describes the decoupling limit of the conifold \GiveonZM. The
mutual locality condition is
\eqn\ml{ F_1\al_2 -F_2\al_1 -\bar F_1 \bar{\al}_2 + \bar F_2
\bar{\al}_1 +(\al_1\al_2-\bar{\al_1}\bar{\al_2})+2(n_1 w_2+n_2
w_1) \in 2\IZ~. }
This condition provides a way to see that in the noncompact case
(in which $w_i=0$), there are no physical states in the $(NS,R)$
and $(R,NS)$ sectors. Indeed, vertex operators in these sectors
are not mutually local with respect to themselves.

When we compactify, we find the theories depicted in figure 1.
Line 1 describes the compactification of type 0 string theory on a
circle. As $R\to\infty$ it approaches the noncompact $0B$ theory;
as $R\to 0$ one finds the noncompact $0A$ theory. For any value of
$R$ there are bulk tachyons in the spectrum.

Other standard compactifications are the super-affine ones,
denoted by lines 2, 3 in figure 1. Since they do not contain
spacetime fermions, these compactifications are simple
generalizations of the super-affine compactification of critical
type 0 string theory on $\IR^{8,1} \times S^1$. For example the
spectrum of line 2 is
 \eqn\linetwo{\eqalign{
 & (0,0,0,0)~,\;\;(0,1,0,1)~, \;\; n\in \IZ~,\;\; w\in 2\IZ~, \cr
 & (1,0,1,0)~,\;\;(1,1,1,1)~, \;\; n\in \frac12 +\IZ~,\;\; w\in 2\IZ~, \cr
 & (1,0,1,1)~,\;\;(1,1,1,0)~, \;\; n\in \IZ~,\;\; w\in 2\IZ +1~,\cr
 & (0,0,0,1)~,\;\;(0,1,0,0)~,\;\; n\in \frac12 + \IZ~,\;\; w\in
 2\IZ+1~.\cr }}
When $R\rightarrow \infty$ we get 0B and when $R\rightarrow 0$
we get 0A that is equivalent to 0B with $R\rightarrow \infty$. Again
the spectrum contains tachyons for any value of $R$.

Another way to interpolate between the noncompact 0A and 0B
theories is to twist by  $(-)^{\al + F}$. The resulting theory,
denoted by line 4 in figure 1, has the spectrum
 \eqn\linefourfour{\eqalign{
 & (0,0,0,0)~,\;\;(1,1,1,1)~, \;\; n=m~,\;\; w=l~,\;\; m-l\in2 \IZ~, \cr
 & (1,0,0,0)~,\;\;(0,1,1,1)~, \;\; n= \frac12 +m~,\;\; w= \frac12 + l~,\;\; m-l\in2
 \IZ~,\cr
 & (0,0,1,0)~,\;\;(1,1,0,1)~, \;\; n= \frac12 +m~,\;\; w= -\frac12 +l~,\;\; m-l\in2 \IZ~,
  \cr
  & (0,1,0,1)~,\;\;(1,0,1,0)~, \;\; n=1+ m~,\;\; w=l~,\;\; m-l\in2
  \IZ~.\cr }}
When $R\rightarrow \infty$ we get 0B. When $R\rightarrow 0$ we get
0B, which is equivalent to 0A with $R\rightarrow \infty$. Thus
this line also interpolates between the noncompact 0A and
noncompact 0B theories.

\ifig\tspace{The moduli space of string theories on $\IR^{3,1}
\times \IR_{\phi} \times S^1$. Line 1 is the usual
compactification that connects  type 0A to type 0B. Lines 2 and 3
are the super-affine compactifications of 0B and 0A. The black
boxes represent the self-dual points along the super-affine lines.
The only point at which the theory is free of bulk tachyons is the
supersymmetric one which is on the line of twisted
compactification (line 4).} {\epsfxsize3.0in\epsfbox{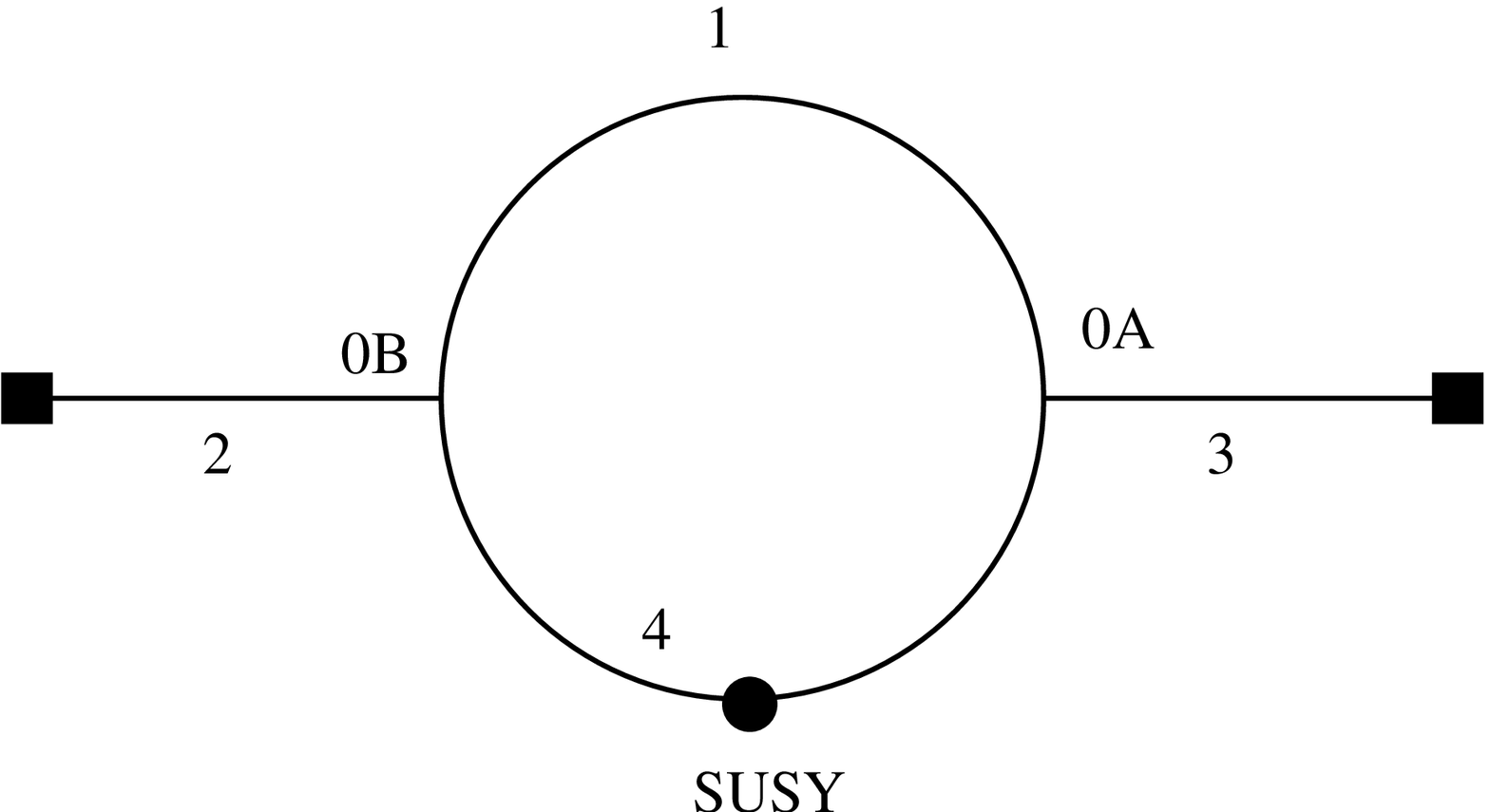}} Starting with
any line in \linefourfour\ we generate the other lines by acting
with $(1,0,0,0) (n=\frac12, w=\frac12)$  and $(0,0,1,0)
(n=\frac12, w=-\frac12)$. For $R=Q=\sqrt{2}$ these operators are
holomorphic and are identified with the supersymmetry generators
\spacetimesup. This is the supersymmetric point of \KutasovUA. As
mentioned in section 2, this is the only theory in the moduli
space depicted in figure 1 which is tachyon free.

\newsec{Six dimensional backgrounds I: stability analysis}

A particular linear dilaton spacetime of the form \aslindil\ is
the CHS background
\CallanAT\
 \eqn\chs{\IR^{5,1}\times\IR_\phi\times SU(2)_k~,} which
describes the near-horizon geometry of $k$ $NS5$-branes. The slope
of the linear dilaton depends on the number of fivebranes via the
relation \eqn\linslopek{Q=\sqrt{2\over k}~.} This background can
also be written in the form \ksnon, where the compact manifold is
$\CM_k=SU(2)_k/U(1)$ (the $N=2$ minimal model with $c=3-{6\over
k}$), and the discrete group is $\Gamma=\IZ_k$. This follows from
the  decomposition of $SU(2)_k$
 \eqn\sucoset{SU(2)_k= \left({SU(2)_k\over U(1)}\times S^1_k\right)/\IZ_k~. }
For $k=2$, the $N=2$ minimal model $\CM_2$ is empty, and the
background \chs\ reduces to \eqn\pa{\IR^{5,1}\times\IR_\phi\times
S^1~.} In the next subsection we study the moduli space
corresponding to this case, and then move on to general $k$.

\subsec{Two fivebranes}

In this case, $F$ takes the values, $(0,1)$ in the NS sector and
$(-{1\over2}, {1\over2})$ in the R sector.  It is still defined
mod 2, however $(-)^F$ belongs to a $\IZ_4$ rather than a $\IZ_2$
group. The mutual locality condition is
\eqn\lu{ F_1\al_2 -F_2\al_1 -\bar F_1 \bar{\al}_2 + \bar F_2
\bar{\al}_1 -\frac12 (\al_1\al_2-\bar{\al_1}\bar{\al_2})+2(n_1
w_2+n_2 w_1) \in 2 \Bbb Z~. }
The two noncompact theories are
 \eqn\fg{\eqalign{
 & 0B: \;\;
 (0,0,0,0)~,\;\;(0,1,0,1)~,\;\;(1,{1\over2},1,{1\over2})~,
 \;\;(1,-{1\over2},1,-{1\over2})~,\cr
 & 0A: \;\;
 (0,0,0,0)~,\;\;(0,1,0,1)~,\;\;(1,{1\over2},1,-{1\over2})~,
 \;\;(1,-{1\over2},1,{1\over2})~.\cr
 }}
Upon compactification,  we find the theories depicted in figure 2.
Lines 1, 2 and 3 are as in the previous section since they do not
contain spacetime fermions. Two additional lines of theories
(lines 4, 5) are obtained by starting with the supersymmetric
theories of \KutasovUA\ (either  IIB or  IIA) and varying $R$. At
the supersymmetric point, the spectrum is determined by the
supercharges, that in the present notation have
\eqn\ue{\eqalign{
& G_1 =(1,{1\over2},0,0)~,\;\;
n={1\over4}~,\;\;w={1\over2}~,\cr & G_2=(0,0,1,{1\over2})~,\;\;
n={1\over4}~,\;\;w=-{1\over2}~,\cr }}
in type IIB,  and
\eqn\upwe{\eqalign{
& G_1 =(1,{1\over2},0,0)~,\;\;n={1\over4}~,\;\;w={1\over2}~,\cr
& G_2=(0,0,1,-{1\over2})~,\;\;
n={1\over4}~,\;\;w=-{1\over2}~, \cr}}
in  type IIA. One can check that $G_1$ and $G_2$ are mutually
local and are holomorphic at $R=Q=1$.

All the states in the theory can be obtained by acting $l_1$ times
with $G_1$ and $l_2$ times with $G_2$ on $(0,0,0,0)$. We denote
these states by $[l_1, l_2]=(l_1, {l_1\over 2}, l_2, \pm {l_2\over
2}),$ with a plus sign in type IIB and a minus sign in type IIA.
$\alpha$ and $F$ are defined modulo two, thus  $0\leq l_1, l_2
\leq 3. $ The momentum and winding modes in each of the sixteen
sectors are determined by the mutual locality of $[l_1, l_2]$ with
respect to $G_1$ and $G_2$. From \lu\ we get
\eqn\uw{n\in \IZ +\frac14 (l_1 +l_2)~,~~~~ w \in 2 \IZ +\frac12
(l_1-l_2)~,~~~~n+{w\over2}\in 2\IZ +{l_1\over2}~. }
The tachyon field, for example, is in the $[2,2]=(0,1,0,1)$
sector. Thus the allowed winding and momentum modes for the
tachyon satisfy
 \eqn\eqss{ n, {w\over2}\in \IZ~, ~~~n+{w\over2}\in 2\IZ +1~.}
The lowest momentum mode has $n=1$, while the lowest winding
mode has $w=2$, in agreement with the discussion of section 2.

Since the spectrum is presented above in terms of $n$ and $w$ (as
opposed to $p_L$ and $p_R$), it is simple to generalize the
discussion to arbitrary radius. Away from the supersymmetric point
$G_1$ and $G_2$ are no longer holomorphic; hence, SUSY is broken.
As explained in section 2,  the theory is tachyon free for
\eqn\notachyon{ \frac34\leq R^2\leq \frac{4}{3}~. } The
supersymmetric point, $R=1$, is  in this range. As explained in
section 2, it is self-dual; it is clear from \chs, that at this
point there is an enhanced  $SU(2)_L\times SU(2)_R$ symmetry. As
$R\to 0,\infty$ we find the noncompact 0B, 0A theories.
\ifig\tspace{The moduli space of string theory in $\IR^{5,1}\times
\IR_{\phi} \times S^1$. Lines 1,2 and 3 are as in figure 1. The
black boxes represent the self-dual points along the super-affine
lines. The thick lines are the regions that are free of bulk
tachyons.} {\epsfxsize3.0in\epsfbox{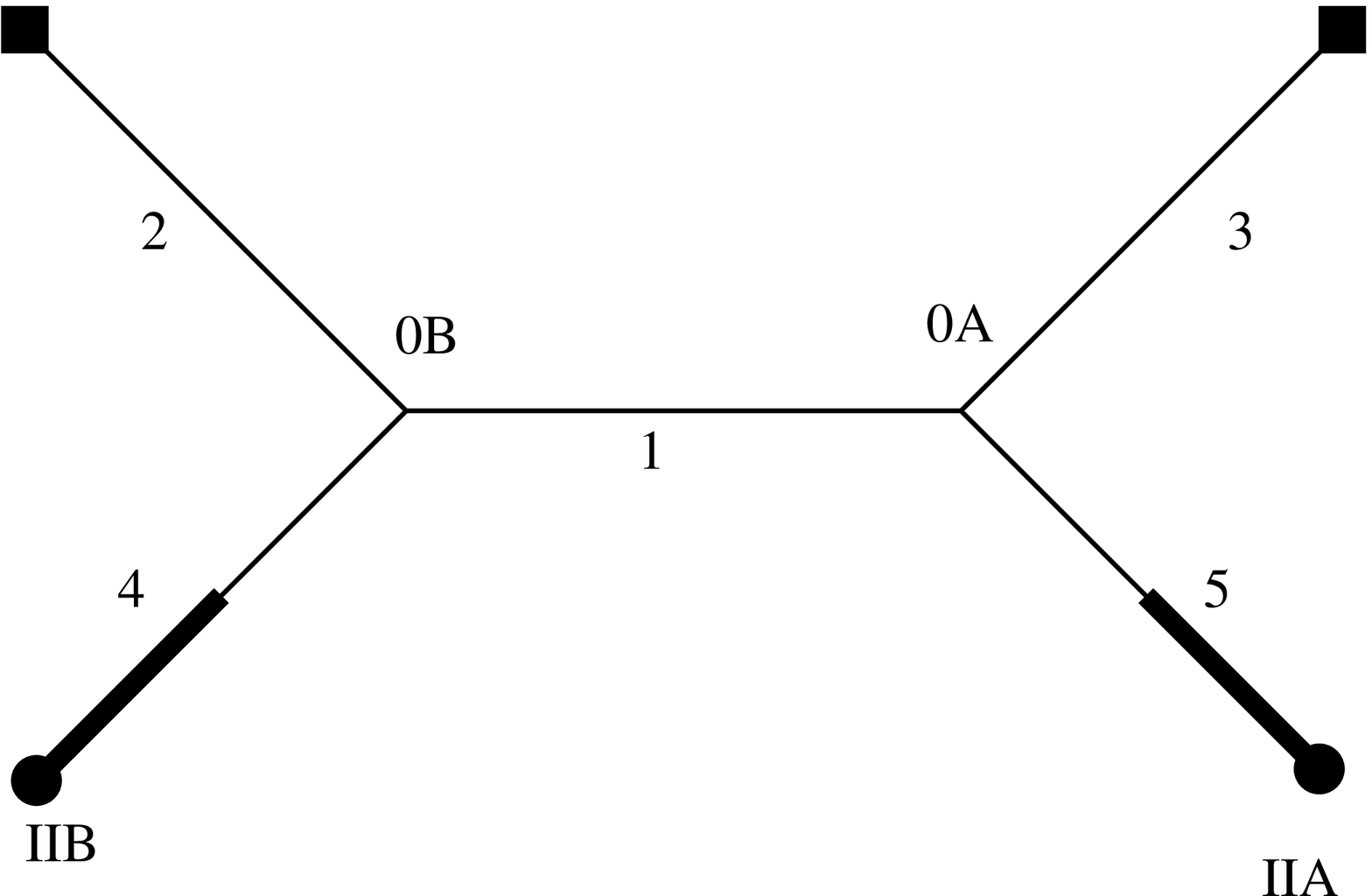}}

\subsec{$k$ fivebranes}

In this subsection we discuss the CHS background \CallanAT\ \chs\
in the presence of the supersymmetry breaking deformation
 \eqn\pert{  \lambda \int d^2 z J_3 \bar{J}_3~,  }
where $J_3, \bar{J}_3$ are particular Cartan subalgebra generators
of $SU(2)_L$ and $ SU(2)_R$. Such a deformation has been
investigated by many authors including
\refs{\HassanGI\KiritsisIU\SuyamaXK-\ForsteKM}. Using the
decomposition \sucoset, this corresponds to changing the radius of
the $S^1$ away from its original value. From the spacetime point
of view, this deformation squashes the three-sphere transverse to
the fivebranes, and breaks the $SO(4)$ symmetry of rotations
around the fivebranes down to $SO(2)\times SO(2)$.

The lightest modes in the background \chs\ are gravitons, whose
vertex operators are given, in the $(-1,-1)$ picture, by\foot{To
simplify the equations we have omitted the contribution of
$\IR^{5,1}\times \IR_\phi$ to these vertex operators.}
 \eqn\gravvert{\left(\psi\bar\psi V_j\right)_{j+1;m,\bar
 m}~.}
The dimension of these operators  is
 \eqn\dimop{\Delta=\half+{j(j+1)\over k}~,}
where $j=0,\half,1,\cdots, {k\over2}-1$. From the point of view of
the decomposition \sucoset\ we can think of \dimop\  as being a
sum of two contributions. One comes from the $SU(2)\over U(1)$
part and is equal to
\eqn\sutwomod{\Delta^{SU(2)\over U(1)}=\half+
{j(j+1)-m^2\over k}~,~~~~~~~m=-j-1,-j,-j+1,\cdots,j+1~,} with a
similar formula for the right-movers. The second contribution to
the dimension comes from the $U(1)$ part and is equal to
 \eqn\uonemod{\Delta^{U(1)}={m^2\over k}~,}
again with a similar formula for the right-movers. The
$U(1)$ part of the dimension comes from momentum and winding
on a circle. Thus, we  have
 \eqn\matchsone{\Delta^{U(1)}=\half
 p_L^2~,\;\;\;\;\; \bar\Delta^{U(1)}=\half p_R^2~, }
with
 \eqn\nnww{p_L={n\over R}+{wR\over2}~,\;\;\;\;\; p_R={n\over
 R}-{wR\over2}~.}
Working in conventions where the winding number is an integer,
we have
 \eqn\Rwn{\eqalign{ R=&Q=\sqrt{2\over k}~,\cr n=&{m+\bar m\over k}~,\cr
 w=&m-\bar m~.\cr }}
Note that while the winding number is an integer, the momentum can
be fractional, $n\in \IZ/k$ due to the $\IZ_k$ orbifold in
\sucoset. States that carry non-trivial $\IZ_k$ charge in
${SU(2)_k\over U(1)}$ have fractional momentum $n$ \Rwn.

Now we change the radius from the supersymmetric point $R_{\rm
susy}=Q=\sqrt{2\over k}$ to some other value, $R$.  We would like
to find  the values of $R$ for which the theory is tachyon free.
As we vary $R$, the operators \gravvert\ retain the property that
$L_0=\bar L_0$, so it is sufficient  to focus on, say, the left
moving part. The dimension of the operator \gravvert\ for generic
$R$ is given by (omitting the $\half$ in \sutwomod):
\eqn\dimRR{{j(j+1)-m^2\over k}+\half p_L^2~.} Adding the Liouville
contribution, we have \eqn\dimliouv{{(j+\half)^2-m^2\over k}+\half
p_L^2- \half(\beta+{Q\over 2})^2~.} The delta-function normalizable
states in the cigar have $\beta=-{Q\over2}+ip$, so the last term
in \dimliouv\ is positive. If the expression \dimliouv\ is
positive, the relevant state is massive, while if it is negative,
it is tachyonic.

We see immediately that for all $|m|\le j$, the states in the
throat are never tachyonic. The only way to get tachyons for any
radius $R$ is to consider the case $|m|=|\bar m|=j+1$, \ie\
$m=\bar m=j+1$, or $m=-\bar m=j+1$. The two are related by T
duality, so it is enough to analyze the first, which corresponds
to pure momentum modes, with $p_L=p_R={2(j+1)\over kR}$. The
condition for this momentum mode to be non-tachyonic is
\eqn\condmass{{(j+\half)^2-(j+1)^2\over k}+\half p_L^2>0} or
\eqn\simpsol{R^2<{2\over k}{(j+1)^2\over j+{3\over4}}~.} Thus
$j=0$ is the first state that becomes  tachyonic. This happens at
$R^2={8\over 3k}$, \ie\ $4\over3$ times the original radius at the
supersymmetric point. Repeating the analysis for winding modes we
conclude that the theory is (bulk) tachyon free for
\eqn\epepep{{3 \over 2k} \leq R^2 \leq {8 \over 3k}~, }
in agreement with the result we got for $k=2$, \notachyon.

\subsec{Localized instabilities}

In the previous subsections we saw that in the six dimensional
case, there is a finite range of radii, for which the bulk theory
in the fivebrane throat is non-tachyonic (in fact, it exhibits a
finite mass gap). This is different from the four dimensional
case, where the gap vanishes, and bulk tachyons appear immediately
as we change the radius. We note in passing that in other four
dimensional linear dilaton backgrounds, where the manifold $\CM$
in \ksnon\ is non-trivial, the gap is finite as well, and the
situation is similar to the six dimensional examples.

In theories with a finite gap in the bulk spectrum, there are
no bulk instabilities even when we break supersymmetry (sufficiently
mildly). A natural question, which we will address in this subsection,
is whether such theories have localized instabilities.

In the six dimensional system \chs\  we understand the low energy
dynamics in terms of the theory on $NS5$-branes. Consider, for
example, the type IIB case.\foot{A similar discussion holds for
IIA fivebranes.} At the supersymmetric point, the low energy
theory on the fivebranes is a six dimensional super Yang-Mills
theory with sixteen supercharges, and gauge group $SU(k)$ (for $k$
fivebranes). This theory has a moduli space parameterized by the
locations of the fivebranes in the transverse $\IR^4$. Denoting
the positions of the fivebranes in the four transverse directions
by the $k\times k$ matrices \eqn\defab{A= x_6 +i x_7~,\qquad
B=x_8+i x_9~,} one can parameterize points in the moduli space of
separated fivebranes by the expectation values of gauge invariant
operators such as ${\rm Tr} A^n$, ${\rm Tr} B^n$, etc.

In the supersymmetric theory, it is well understood
\refs{\AharonyUB,\GiveonPX,\GiveonTQ} how to describe
separated fivebranes in the geometry \chs. For example,
if the fivebrane configuration has a non-zero expectation
value of ${\rm Tr} B^{2j+2}$, the worldsheet theory should
be deformed by the operator
\eqn\moduB{e^{-\varphi-\bar\varphi}\psi^+\bar\psi^+V_{j;j,j}
e^{-Q(j+1)\phi}~.}
This operator corresponds to a wavefunction that is supported
in the strong coupling region of the linear dilaton space $\IR_\phi$.
It is an example of a localized state in the throat. This is in
agreement with the spacetime picture, according to which it describes
a scalar living on the fivebranes.

A particularly symmetric deformation of the kind \moduB\ that has
been extensively studied corresponds to placing the fivebranes on
a circle of size $r_0$ in the $B$ plane \eqn\symspread{B_l=
r_0e^{2\pi il\over k},\qquad A_l=0~, ~~~~l=1,2,...,k~.} In the
bulk, this is described by condensation of the mode \moduB\ with
$j={k\over2}-1$. It is useful to write this operator in the
decomposition \sucoset. By looking at equation \sutwomod, and
using the fact that it has $m=j+1={k\over2}$, we see that from the
point of view of $SU(2)\over U(1)$ it is proportional to the
identity operator. In fact, it is the $N=2$ Liouville
superpotential \eqn\ntwoliou{W=e^{i{1\over R}x -{1\over Q}\phi}~,}
where $R=Q$ is the radius at the supersymmetric point \Rwn. As is
well known, the $N=2$ Liouville operator is truly marginal, and
signals the change from $\IR_\phi\times S^1$ to the cigar (here we
are describing it in terms of the T-dual variables, so the N=2
Liouville perturbation is a momentum mode). Note that we are not
assuming here that the string coupling is weak everywhere. It
could be that the string coupling at the tip of the cigar is very
large, but we are studying the weakly coupled region where it is
small.

Now suppose we change the radius $R$ from its original value \Rwn.
There are two different cases to discuss. If we make $R$ smaller,
the operator \ntwoliou, which had dimension one before, will
have dimension larger than one. This means that it is massive,
and we have to dress \ntwoliou\ with a time-dependent part, like
$\cos(Et)$ with a real energy $E$. This means that  $\langle{\rm Tr}
B^k\rangle$, and the size of the circle on which the fivebranes are
placed, $r_0$ \symspread, will oscillate between two extreme values.

On the other hand, if we increase the radius $R$, the dimension of
\ntwoliou\ drops below one, it becomes tachyonic, and to make it
physical we have to dress it with, \eg, $\cosh(Et)$ with a real
$E$. This means that $r_0$ increases without bound (in this
approximation) as $|t|\to\infty$. In the next section we explore
in details  these dynamical processes using various techniques.

\newsec{Six dimensional backgrounds II: Time dependence, field theory and gravity}

In this section we further discuss the deformed $NS5$-brane background with
broken supersymmetry obtained by applying the perturbation \pert. In the first
subsection we construct time-dependent solutions that correspond to rolling
fivebranes. In the second subsection we comment on the description of the
deformed fivebrane system using the low energy field theory and supergravity.
The latter is valid at large $k$.

\subsec{Rolling fivebranes}

At the end of the last section we briefly mentioned the time-dependent
solutions obtained by displacing the fivebranes from the origin of the
moduli space $A=B=0$ \defab\ in the presence of the supersymmetry
breaking deformation \pert. We only discussed what happens to leading
order in the separation of the fivebranes \ntwoliou, since we only imposed
the leading order condition that the dressed perturbation be marginal. In
general, one expects the solution to receive corrections, and for
the solutions mentioned in the previous section we expect these
to be large. The reason is that these solutions correspond to
accelerating fivebranes, and are expected to radiate. Since the
tension of the $NS5$-branes scales like $1/g_s^2$, the closed strings
radiation (that goes like $G_N T_5$) is an order one effect that
influences the solution in a non-trivial way at tree
level.\foot{This is different from the well studied case of
unstable D-branes with a rolling tachyon \SenNU. There, the closed
strings radiation is an order $g_s$ effect
\refs{\LambertZR,\GaiottoRM}, and so can be neglected (at least
formally) in the leading approximation.}

Nevertheless, it is possible to find exact solutions of the
classical string equations of motion that correspond to rolling
fivebranes. Consider first the case of decreasing the radius $R$,
such that the $N=2$ Liouville operator \ntwoliou\ becomes
irrelevant (or massive). In that case, we can make it marginal by
replacing \ntwoliou\ with \eqn\timedep{W= e^{{i\over
R}x+i\omega_Rt-{1\over Q}\phi}~.} The mass-shell condition
requires that \eqn\mmass{\omega_R^2={1\over R^2}-{1\over Q^2}~.} We
can define a coordinate $x_{\rm new}$ by \eqn\ynewdef{{1\over
Q}x_{\rm new}={1\over R}x+\omega_Rt~.} The condition \mmass\
implies that $x_{\rm new}$ is canonically normalized (when $x$ and
$t$ are). In terms of this new field, the perturbation separating
the fivebranes \timedep\ looks precisely like the $N=2$ Liouville
perturbation again. Thus, we conclude that \timedep\ is a truly
marginal deformation. It describes the fivebranes placed on the
slope of the quadratic potential  and rotating around the circle
with angular velocity $\omega_R$ (see figure 3a). The existence of
an exact CFT description implies that there should also be a
supergravity solution associated with that system. This solution
is described in the appendix.

\ifig\tspace{Time dependent processes which have exact  CFT
descriptions:  (a) Fivebranes rotating around a circle with a
constant angular velocity; (b) Fivebranes moving radially with a
constant radial velocity.} {\epsfxsize4.5in\epsfbox{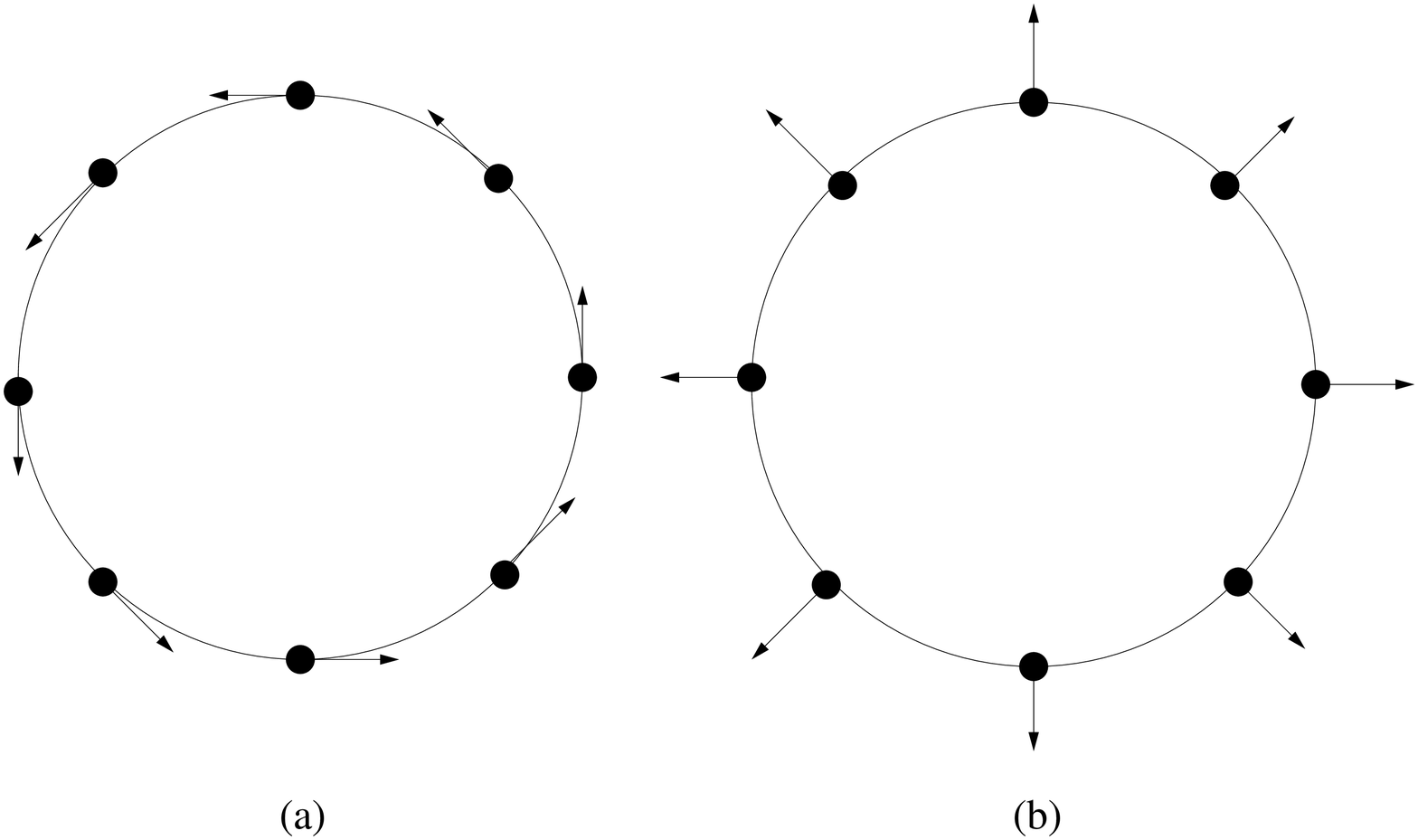}}

The orthogonal combination of $(t,x)$ is timelike,
\eqn\tnew{{1\over Q} t_{\rm new}={1\over R} t+\omega_R x~.}
The original periodicity of $x$, $x\sim x+2\pi R$ translates in
terms of the new coordinates to:
\eqn\newperiod{(t_{\rm new}, x_{\rm new})\sim
(t_{\rm new}, x_{\rm new}) +2\pi Q(\omega_RR,1)~.}

\noindent One can describe the solution of fivebranes rotating
around the circle by a coset CFT. The supersymmetric solution
corresponding to fivebranes on a circle can be thought of as
\eqn\coscft{\IR^{5,1}\times SL(2,\IR)_k\times SU(2)_k} modded out
by the null $U(1)$ symmetry generated by the current
\eqn\nulluone{J=J_3-K_3~,} where $J_3$ and $K_3$ are the Cartan
subalgebra generators of  $SL(2)$ and  $SU(2)$ respectively. In
order to go from this solution to the one described by \timedep,
we need to mix the $U(1)$ current $K_3$ with the timelike $U(1)$,
$i\partial t$. Ignoring the periodicity conditions, this is simply
a boost transformation. Thus, in terms of the coset description
\coscft, \nulluone, we want to gauge the current
\eqn\modnulluone{J_3-\beta_1K_3-{i\beta_2\over Q}\partial t~,}
where the boost parameters $\beta_i$ satisfy
$\beta_1^2-\beta_2^2=1$. This provides an exact CFT description of
the system of fivebranes rotating around the circle.

Naively, one might expect that as the fivebranes are rotating
around the circle they should radiate, and the solution should be
modified, already at tree level. This seems in contradiction to
the fact that we found an exact solution of the full equations of
motion of classical closed string theory. The resolution of
this puzzle is that in the near-horizon limit, the solution with
fivebranes rotating around the circle corresponds to motion with
constant velocity, and not constant acceleration (since the
near-horizon metric for the angular $S^3$ is $d\Omega^2$ and not
$r^2d\Omega^2$). Thus, it does not radiate.

This should be contrasted with the case where we place the
fivebranes at some distance from the origin and let them go. In
this case, one expects the system to oscillate about the minimum,
with  $r_0\sim\cos\omega_Rt$. This amounts to replacing
$\exp(i\omega_Rt)$ in \timedep\ with $\cos(\omega_Rt)$. From the
spacetime point of view, we expect non-zero radiation in this
case, since this solution corresponds to a time dependence of
$\phi$ of the form $\phi\sim\log\cos\omega_Rt$. Thus, there is in
this case non-zero acceleration, and we expect the solution to be
modified at the tree level. This is natural from the worldsheet
point of view. If we replace $\exp(i\omega_Rt)$ by
$\cos(\omega_Rt)$ in \timedep, the operator \timedep\ ceases to be
holomorphic, and one expects it to become marginally relevant
(like the marginal Sine-Gordon and non-Abelian Thirring model
couplings). Presumably, this behavior of the worldsheet theory is
directly related to the classical radiation that one expects in
spacetime.

For the other case, of increasing\foot{T-duality relates
increasing $R$ to decreasing $R$, while exchanging momentum and
winding modes.  Therefore, instead of increasing $R$ we can
continue to decrease it and study other classical solutions.}
radius $R$, for which $w_R$ is imaginary, the operator \ntwoliou\
is relevant, and we expect to have to dress it with a real
exponential in time. Thus, instead of \timedep\ we now have
\eqn\tachh{W= e^{{i\over R}x+w_Rt-{1\over Q}\phi}~,} and $w_R$ is
given by (similarly to \mmass) \eqn\wwrr{w_R^2={1\over
Q^2}-{1\over R^2}~.} We would like again to make an $N=2$ Liouville
perturbation out of \tachh, so we define a new Liouville
coordinate $\phi_{\rm new}$ by \eqn\defphi{{1\over R}\phi_{\rm
new}={1\over Q}\phi-w_Rt ~.} Again, due to \wwrr, $\phi_{\rm new}$
is canonically normalized. Now, in terms of $\phi_{\rm new}$ and
$x$ we again have an $N=2$ Liouville theory, so this is an exact
solution of the equations of motion of the theory, corresponding
to  fivebranes approaching the origin of the $B$ plane as
$t\to-\infty$, and moving out to infinity  at $t\to+\infty$ (see
figure 3b). This is an analog of the ``half-brane''  from unstable
D-brane dynamics. Note that, as in the discussion of fivebranes
rotating around a circle, during the time evolution no radiation
is emitted, since the fivebranes are moving with constant velocity
in the $\phi$ direction. The relevant ``rolling fivebrane''
supergravity solution is constructed in the appendix.

One thing that is different in this case relative to the previous
one is that since we are mixing the time coordinate with $\phi$ to
make the cigar, the remaining combination of $\phi$ and $t$,
\eqn\ttnneeww{{1\over R}t_{\rm new}={1\over Q}t-\omega_R\phi}
which does not participate in the cigar and is timelike, has a
linear dilaton associated with it. More precisely, one finds,
\eqn\qnew{Q_{\phi_{\rm new}}=R~,\;\;\;Q_{t_{\rm new}}=QR\omega_R~.}
This signals the fact that at early times ($t,t_{\rm
new}\to-\infty$), when the fivebranes are on top of each other, we
have a strong coupling problem, despite the fact that we have
constructed a cigar which avoids the strong coupling problem in
the direction $\phi_{\rm new}$.

Again, one can describe the time-dependent background \tachh\ by
an exact coset CFT. As we see from \defphi, \ttnneeww, in this
case we need to perform a boost mixing $\phi$ and $t$, or $J_3$ and
$\partial t$. One can describe this background as \coscft\ modded
out by the null current \eqn\newnull{J_3+{\beta_1\over Q}\partial
t-\beta_2 K_3~,} with $\beta_1^2+\beta_2^2=1$.

The solution describing fivebranes running to infinity as a
function of time is an example of localized tachyon condensation.
In our case we are able to find an exact CFT that describes this
time-dependent process, and smoothing out of the singularity,
while in the case of non-supersymmetric orbifolds \AdamsSV\ no
exact time-dependent solutions of this sort were found. The reason
for the difference is that we took the near-horizon limit of a
throat, in which, as we have seen, the smoothing out of the
singularity proceeds via a process in which no radiation is
emitted, whereas the analysis of \AdamsSV, takes place in the full
geometry where one expects the solution to be much more
complicated and to involve non-trivial radiation effects.\foot{The
relation between localized tachyons on non-supersymmetric
orbifolds and throat theories was discussed in \HarveyWM.}

A related point is that here we wrote  an exact solution which is
an analog of a half S-brane \refs{\GutperleAI,\SenNU} for decaying
D-branes. There should exist a more complicated solution where the
distance between the fivebranes starts at $t\to-\infty$ very
large, decreases to a minimal value, and then increases back to
infinity. This solution would be an analog of the full S-brane. In
this case the $\phi$ coordinate accelerates near the turning
point, and one expects non-trivial radiation effects at tree
level.

\subsec{Dual field theory and supergravity }

In this subsection we consider the deformation \pert\ from the point
of view of the dual field theory and supergravity descriptions. We focus
on the type IIB case. In this case the dual field theory is a six dimensional
SYM theory with sixteen superchrages and an $SU(k)$ gauge group.
This field theory is infrared free and is believed  to
resolve the strong coupling singularity associated with the CHS
background  \ItzhakiDD. Since the wave function of \moduB\ is
concentrated in the strong coupling region, one can study the relevant
dynamics using  the dual field theory description.

The relationship between the bulk modes in the CHS background and
chiral field theory operators was described in \AharonyUB. The
vertex operators \gravvert\ correspond to symmetric, traceless
combinations of the four scalars \defab.  In particular, the
deformation \pert\ is dual to a mass term for the scalars \eqn\rt{
\alpha \Tr (A^* A - B^* B)~.} Thus, with the deformation two of
the four flat directions become massive, while the other two
become tachyonic. This fits nicely with the results of the
previous subsection. The relation between $\alpha$ and $R$ is the
following. Normalizing the field $B$ such that the Lagrangian
takes the form
 \eqn\lagbb{\CL=|\dot B|^2+\alpha|B|^2~,}
and comparing the solution of the equation of motion
$B=r_0 e^{\pm i\sqrt{\alpha}t}$ to \timedep, we find that
\eqn\valalpha{\alpha={1\over Q^2}-{1\over R^2}~.}

Another way to study the dynamics is to use the supergravity
description, which is accurate for large $k$. First we have to
find the deformation of the CHS background associated with \pert.
In the appendix we describe the relevant solution from the coset
point of view. Here we describe this solution directly in the CHS
language.

The  starting point is the  CHS background associated with $k$
$NS5$-branes \CallanAT,
 \eqn\sol{\eqalign{
  &ds^2=dx_{||}^2+d\phi^2+
 2k\left( d\theta^2+\sin^2\theta d\phi^2_1+ \cos^2\theta
 d\phi^2_2\right)~,\cr
 & B=k(1+ \cos(2\theta)) d\phi_1\wedge d\phi_2~,~~~~~~~
 g_s^2=\exp(-\sqrt{2/k}\phi)~,}}
where $0\leq \theta \leq \pi/2,~0\leq \phi_1, ~\phi_2 \leq 2\pi$.
The transverse coordinates \defab\ are given by
\eqn\rel{ A=\sqrt{2k}~ \exp(\phi \sqrt{1/2k})\sin(\theta) \exp(i
\phi_1)~,~~~~~
B=\sqrt{2k}~ \exp(\phi \sqrt{1/2k})\cos(\theta) \exp(i \phi_2)~.}
This relation can be derived by comparing the field theory and the
supergravity expressions for  the energy of a BPS $D1$-brane (or
of a fundamental string in the near-horizon geometry of $k$ $D5$-branes).

To find the solution associated with the perturbation \pert\ we
recall that this deformation can be viewed
\refs{\HassanMQ,\HassanGI,\GiveonPH} as an $SL(2, \IR)$
transformation  of the usual CHS background. The relevant
$SL(2,\IR)$ transformation is \eqn\sl{ \tau\rightarrow
\tau^{'}=\frac{\tau}{1+\alpha\tau}~,} where $\tau$ is the Kahler
parameter, $ \tau\equiv B_{12}+i\sqrt{g}=
2k\cos(\theta)e^{i\theta}.$ A short calculation yields
 \eqn\newsol{\eqalign{
 & ds^2=dx_{||}^2+d\phi^2+ 2 k\left(
 d\theta^2+\frac{1}{(1+k \alpha )^2+\tan^2\theta}(\tan^2
 \theta d\phi^2_1+ d\phi^2_2) \right)~, \cr
 & B= \frac{2k(1+k\alpha)}
 {(1+k \alpha )^2+\tan^2\theta} d\phi_1\wedge d\phi_2~,~~~~~
 g_s^2=e^{-\phi\sqrt{2/k}}\frac{1+\tan^2\theta}{(1+k \alpha
 )^2+\tan^2\theta}~.}}
This solution is not quite what we are after since it has a conical
singularity at $\theta=0$. This can be fixed by rescaling
$\phi_1 \rightarrow \phi_1/(1+k\alpha)$, which yields
 \eqn\finsol{\eqalign{
 &ds^2=dx_{||}^2+d\phi^2+ 2 k\left(
 d\theta^2+\frac{1}{L^2+\tan^2\theta}(L^2\tan^2
 \theta d\phi^2_1+ d\phi^2_2) \right)~, \cr
 &  B= \frac{2kL^2}
 {L^2+\tan^2\theta} d\phi_1\wedge d\phi_2~,~~~~~
 g_s^2=e^{-\phi\sqrt{2/k}}\frac{1+\tan^2\theta}{L^2+\tan^2\theta}~,}}
where $L=1+\alpha k$. The curvature in string units is small as
long as  $k\gg 1$ and $ L-1\ll 1$.

We wish to show that this background indeed exhibits the same
physics as \rt. Namely, it has two massive directions and two
tachyonic ones. To this end we calculate  the potential felt by a
probe fivebrane that is localized in $\phi$ and on the sphere.
This potential is the strong coupling dual  of \rt. In general
such a calculation need not yield the same result. However, since
for $|\alpha | \ll 1$ we are in the near BPS limit we expect to
find the same potential to leading order in $\alpha$.

To see that this is indeed the case, consider the DBI action of a
probe $D5$-brane propagating in the S-dual background,
\eqn\finsolll{\eqalign{
 &ds^2=g_s\left[ dx_{||}^2+d\phi^2+ 2 k\left(
 d\theta^2+\frac{1}{L^2+\tan^2\theta}(L^2\tan^2 \theta d\phi^2_1+
 d\phi^2_2) \right) \right]~, \cr
 &F_3= \frac{-4kL^2\tan(\theta)(1+\tan^2\theta)}
 {(L^2+\tan^2\theta)^2} d\phi_1\wedge d\phi_2 \wedge d\theta~,~~~~~
 g_s^2=e^{\phi\sqrt{2/k}}\frac{L^2+\tan^2\theta}{1+\tan^2\theta}~.}}
The six-form potential that couples to the $D5$-brane probe is $
A_{||}=Le^{\phi\sqrt{2/k}}$ and  the DBI potential reads
\eqn\pot{V=g_s^{-1}\sqrt{g_{||}}-A_{||}=e^{\phi\sqrt{2/k}}\left(
\frac{L^2+\tan^2\theta}
{1+ \tan^2\theta}-L\right).}
Using \rel\ we can express this potential in terms of the field theory
variables. For small deformation, $\alpha \ll 1$, we find,
\eqn\potsugra{V=\alpha\left(A^*A-B^*B\right)~, }
in agreement with \rt\ .

\newsec{Two dimensional background}

In this section we study superstrings on $\IR^{1,1} \times \IR_{\phi}
\times S^1$, with $Q=\sqrt{3}$. The analysis here is fairly similar
to the one in section 4.1. Like in the six dimensional case, $F$ takes
the values $(0,1)$ in the NS sector and $(-{1\over2}, {1\over2})$ in the
R sector. The mutual locality condition is slightly different than
in the six dimensional case
\eqn\mltwo{ F_1\al_2 -F_2\al_1 -\bar F_1 \bar{\al}_2 + \bar F_2
\bar{\al}_1 +\frac12 (\al_1\al_2-\bar{\al_1}\bar{\al_2})+2(n_1
w_2+n_2 w_1) \in 2\IZ~. }
The moduli space is very similar to that of figure 2.
The type 0 and super-affine lines $1,2,3$ are as there. As in the
six dimensional case, there are two additional lines of theories
that are obtained by starting with the supersymmetric theories
(either IIA or IIB) and varying $R$. These theories can be
analyzed using the same approach as in section 4.1. Now the
supersymmetry generators are
 \eqn\kl{\eqalign{
 & G_1 =(1,{1\over2},0,0)~,\;\;
 n=\frac34~,\;\;w=\frac12~,\cr & G_2=(0,0,1,{1\over2})~,\;\;
 n=\frac34~,\;\;w=-\frac12~,\cr}}
in type IIB, and
 \eqn\yv{\eqalign{
 & G_1 =(1,{1\over2},0,0)~,\;\;
 n=\frac34~,\;\;w=\frac12~,\cr & G_2=(0,0,1,-{1\over2})~,\;\;
 n=\frac34~,\;\;w=-\frac12~,\cr }}
in type IIA. With the help of \mltwo\ we find that the winding and
momentum modes in the $[l_1, l_2]$ sector satisfy
\eqn\pepepe{ n\in \IZ -\frac14 (l_1 +l_2)~,~~~~3w\in 2\IZ +\frac12 (l_2- l_1)~,~~~~
 n +\frac32 w \in 2 \IZ -\frac12 l_1~.  }
For the tachyon modes (in the  $[2,2]$ sector), we find, in
agreement with section 2, that the lowest momentum is one, and the
lowest winding is $2\over3$. As explained in section 2 this
implies that the theory is free of bulk tachyons for $\frac32 \leq R
\leq 2$. The supersymmetric radius, $R=\sqrt{3}$, is self-dual under
T-duality. However, here, unlike the six dimensional case, the
symmetry is not enhanced to $ SU(2)_L \times SU(2)_R$ at this point.
Again, these lines interpolate between 0B (0A) and 0B (0A).

While the physics in the bulk of the linear dilaton throat is
quite similar to the six dimensional case, the localized dynamics
is different. In the six dimensional case, there was an
instability due to localized tachyons such as the $N=2$ Liouville
mode \ntwoliou, which was massless for $R=Q$ and became tachyonic
when $R\neq Q$. This deformation exists also in the two
dimensional case. However since  now $Q>\sqrt{2}$, this mode is
non-normalizable (for recent discussions, see
\refs{\KarczmarekBW,\GiveonMI} ), and hence it cannot dynamically
condense. In fact, in this case there are {\it no} normalizable
localized tachyons, so the non-supersymmetric model is locally
stable.

\newsec{Three dimensional backgrounds}

The  theories we considered in the previous sections have the
property that the $N=2$ Liouville operator \ntwoliou\  is in the
spectrum. Since this mode is tachyonic for $R\not=Q$,
$Q<\sqrt{2}$, one might be tempted to conclude that there are no
stable non-supersymmetric deformations of linear dilaton backgrounds
with a small dilaton slope. This conclusion is incorrect.

A counter example is the background
\eqn\mh{\IR^{2,1}\times \IR_\phi\times SU(2)_{k_1}\times
SU(2)_{k_2}~, }
with
\eqn\defqq{Q=\sqrt{2\over k}~, \qquad {1\over k}={1\over
 k_1}+{1\over k_2}~.}
We will see that a small non-supersymmetric deformation of the form \pert,
acting on either of the two $SU(2)$'s in \mh, does not lead to instabilities
in this case.

The background \mh\ is the near-horizon geometry of the
intersection of $k_1$ coincident $NS5$-branes stretched in the
directions $(012345)$ and $k_2$ coincident $NS5$-branes stretched
in $(016789)$ (see \ItzhakiTU\ for a recent discussion). This
brane configuration is Poincare invariant in $1+1$ dimensions, but
the near-horizon geometry \mh\ exhibits a higher Poincare
symmetry, in $2+1$ dimensions \ItzhakiTU .

In section 4 we considered the system of $k$ parallel $NS5$-branes
and argued in three different ways for instability when \pert\
is turned on. For the background \mh\ all these arguments indicate
that the system is stable.

\item{(1)}  From the worldsheet point of view the instability was
due to the fact that the $N=2$ Liouville mode became tachyonic when
we changed the radius. Now, however, this mode is not in the spectrum
\ItzhakiTU . This is possible because this theory is not of the form
\ksnon.

\item{(2)}  The dual field theory argument for the instability was
that some of the flat directions at $R=R_{\rm susy}$ became
tachyonic when $R\neq R_{\rm susy}$. The field theory dual to \mh
, however, has a mass gap \refs{\ItzhakiTU,\LinNH}. Hence a small
deformation of the parameters cannot lead to a tachyonic mode. The
flat directions in the case of $k$ parallel $NS5$-branes were
associated with moving the branes in the transverse directions. In
our case these deformations are massive, since each stack of
fivebranes is wrapped around a three-sphere (see \ItzhakiTU\ for a
more extensive discussion).

\item{(3)} A complementary argument  for instability came from the probe
fivebrane dynamics (see section 5.2). Is it possible that a probe fivebrane
in the background \mh\ that respects three dimensional Poincare
invariance has a flat direction and  can become unstable away from
the supersymmetric point? To verify that this is not the case let
us calculate the potential experienced by such a probe fivebrane.
Again it is convenient to work with S-dual variables and study the
DBI action for a probe $D5$-brane. The S-dual metric and dilaton
take the form
 \eqn\sduals{ds^2=g_s\left[-dx_0^2+dx_1^2+dx_2^2+d\phi^2+2k_1 d\Omega_3^2+
 2k_2 d\tilde{\Omega}_3^2\right]~,~~~~g_s^2=e^{\phi Q}~,}
and the RR-fields are
 \eqn\rr{ F_3= 2k_1\sin(2\theta)d\phi_1 \wedge d\phi_2 \wedge
 d\theta +2k_2\sin(2\tilde{\theta})d\tilde{\phi}_1 \wedge d\tilde{\phi}_2 \wedge
 d\tilde{\theta}~.}
To compute the DBI action of a probe $D5$-brane we need the
dual field strength
 \eqn\eqFs{ \eqalign{
  F_7= \star F_3= 2e^{Q\phi} & \left[ k_1 \sin(2\tilde{\theta})
 (k_2/k_1)^{3/2} dx_{||}\wedge d\phi\wedge d\tilde{\Omega} +\right. \cr
  \qquad & \left.+   k_2 \sin(2\theta)
 (k_1/k_2)^{3/2} dx_{||}\wedge d\phi\wedge d\Omega \right]~,}}
from which we  find the six-form that couples to the D5-brane
 \eqn\sif{A_6=2e^{Q\phi} \left[\frac{k_1}{Q}(k_2/k_1)^{3/2}\sin(2 \tilde{\theta})
 dx_{||}\wedge d\tilde{\Omega}+ \frac{k_2}{Q}(k_1/k_2)^{3/2}\sin(2 \theta)
 dx_{||}\wedge d\Omega \right]~. }
Thus  the potential felt by a probe $D5$-brane stretched along $(x_0, x_1,
x_2, \tilde{S}^3)$ and  localized at some $\phi$,  is attractive
 \eqn\dbip{\int_{\tilde{S}_2}g_s^{-1}\sqrt{g_{||}g_{\tilde{S^2}}}-A_6
 =4\pi^2 V_{||}e^{Q\phi}
\sqrt{2} k_2^{3/2}\left( 1-\sqrt{\frac{k_2}{k_1+k_2}}\right)~.}
The analog of \rel\ for this case implies that the scalar fields are
proportional to $\exp(Q \phi /2)$. Thus  \dbip\ takes the form of
a mass term for the scalars. This is in agreement with the field
theory analysis of \ItzhakiTU\ that leads to a mass gap in the
dual field theory.

\bigskip
 \centerline{\bf Acknowledgements}
We thank J.~Maldacena for a discussion.  DK thanks the Weizmann
Institute, Rutgers NHETC and Aspen Center for Physics for
hospitality during parts of this work. NI thanks the Enrico Fermi
Institute at the University of Chicago for hospitality. The work
of DK is supported in part by DOE grant DE-FG02-90ER40560; that of
NS by DOE grant DE-FG02-90ER40542. NI is partially supported by
the National Science Foundation under Grant No.\ PHY 9802484. Any
opinions, findings, and conclusions or recommendations expressed
in this material are those of the authors and do not necessarily
reflect the views of the National Science Foundation.

\appendix{A}{Time dependent supergravity solutions}

In section 5 we argued that there are  exact CFT's, \timedep\ and
\tachh, that describe certain time dependent $NS5$-brane
configurations. The aim of this appendix is to construct the
supergravity solutions associated with these CFT's. Both \timedep\
and \tachh\ are exactly marginal deformations of the coset $\left(
S^1_k\times \frac{SU(2)_k}{U(1)}\right) /\Bbb Z_k$. When thinking
about the relevant  supergravity solutions it is more natural to
use the $SU(2)_k$ description. Therefore,  we first review the
details of the transformation that takes the $SU(2)_k$ to $\left(
S^1_k\times \frac{SU(2)_k}{U(1)}\right) /\Bbb Z_k$. We start on
the $SU(2)_k$ side (in this appendix we set $\alpha'=1$)
 \eqn\su{
  ds^2=k\left( d\theta^2+\sin^2\theta d\phi^2_1+ \cos^2\theta
 d\phi^2_2\right)~,~~~~~B_{12}=k \cos^2\theta~,}
where $0\leq \theta \leq \pi/2,~0\leq \phi_1, ~\phi_2 \leq 2\pi$.
The complex and Kahler structures associated with this background
are
 \eqn\comc{
 \tau=\frac{g_{12}}{g_{22}}+i\frac{\sqrt{g}}{g_{22}}=i\tan\theta~,~~~~~
 \tau_k=B_{12}+i\sqrt{g}=k\cos\theta e^{i\theta}~.}
To transform to the $\left( S^1_k\times
\frac{SU(2)_k}{U(1)}\right) /\Bbb Z_k$ description  we first apply
T-duality in the $\phi_2$ direction. This amounts to $\tau
\leftrightarrow \tau_k$ (for a review see \GiveonFU ) so we find
the T-dual background to be
 \eqn\bacone{
 ds^2=k(d\theta^2+d\phi_1^2)+2d\phi_1d\phi_2+\frac{1}{k\cos^2\theta}d\phi_2^2~,
 ~~~~B_{12}=0~.}
Defining new coordinates
 \eqn\trans{\tilde{\phi_1}=\phi_1+\phi_2/k~,~~~~\tilde{\phi_2}=\phi_2/k~,}
we find
 \eqn\baco{ ds^2=k\left( d\theta^2+ d\tilde{\phi}_1^2+\tan^2\theta
 d\tilde{\phi}_2^2\right).}
Note that now we have a $\Bbb Z_k$ identification,
$(\tilde{\phi}_1, \tilde{\phi}_2) \sim (\tilde{\phi}_1+2\pi/k,
\tilde{\phi}_2+2\pi/k)$, that implies that the background is
indeed $\left( S^1_k\times \frac{SU(2)_k}{U(1)}\right) /\Bbb Z_k$.
As usual the dilaton picks an extra factor from the T-duality
transformation
 $g_s\rightarrow g_s ( \det g_{\rm new}/\det
 g_{\rm old})^{1/4},$ so we have
 \eqn\dil{ g_s\to \frac{g_s}{\cos\theta}~.}
Now we shall deform the $\left( S^1_k\times
\frac{SU(2)_k}{U(1)}\right) /\Bbb Z_k$ side in various ways and see
what these deformations give on the $SU(2)_k$ side.

\item{(1)} The supersymmetry breaking deformation \pert\ corresponds
to changing the radius of the $S_1$ from $\sqrt{k}$ to $L \sqrt{k }$.
The resulting metric on the squashed sphere is
\eqn\bacoo{ ds^2=k\left( d\theta^2+L^2 d\tilde{\phi}_1^2+\tan^2\theta
d\tilde{\phi}_2^2\right)~,~~~~~g_s=\frac{1}{\cos\theta}~.}
The change of coordinates \trans\ leads to
\eqn\eqone{ds^2=k\left( d\theta^2 + L^2 d\phi_1^2+\frac{2L^2}{k} d\phi_1
d\phi_2 +\frac{L^2+\tan ^2\theta}{k^2}d\phi_2^2\right)~.}
Applying T-duality in the $\phi_2$ direction we get
\eqn\newsol{\eqalign{
&ds^2=k\left( d\theta^2 +\frac{ L^2 \tan^2\theta}{L^2+\tan^2
\theta} d\phi_1^2+ \frac{1}{L^2+\tan^2\theta} d\phi_2^2\right) \cr
&B_{12}=\frac{kL^2}{L^2+\tan^2 \theta}~,
~~~~~g_s^2=\frac{1+\tan^2\theta}{L^2+\tan^2\theta}~.}}
This agrees with the $SU(2)$ contribution to \finsol.

\item{(2)} Another useful background is obtained by adding a
direction, $\rho$, and combining it with $\tilde{\phi}_1$ to
form a cigar $\left(\frac{SL(2)_k}{U(1)} \times
\frac{SU(2)_k}{U(1)}\right)/\Bbb Z_k$. This does not change
the asymptotic radius of the $S^1$, so at large $\rho$ the
background is $SU(2)_k \times \IR $. The metric is
\eqn\slsu{ds^2=k\left( d\theta^2 +\tan ^2\theta
d\tilde{\phi}_2^2+d\rho^2+\tanh ^2\rho
d\tilde{\phi}_1^2\right)~,~~~g_s=\frac{1}{\cos\theta \cosh \rho}~,}
where $(\tilde{\phi}_1, \tilde{\phi}_2) \sim
(\tilde{\phi}_1+2\pi/k, \tilde{\phi}_2+2\pi/k)$. Now we wish to find
the effect of this deformation in the $SU(2)$ variables. Using \trans\
we get
\eqn\slsut{ds^2=k\left(d\theta^2+d\rho^2+\tanh ^2\rho
d\phi_1^2\right) + 2\tanh ^2 \rho d\phi_1 d\phi_2 +\frac{1}{k}(\tan
^2\theta
 +\tanh ^2\rho ) d\phi_2^2~.}
Applying T-duality we get
 \eqn\susl{\eqalign{
 & ds^2=k\left( d\theta^2+d\rho^2+\frac{\tan^2\theta \tanh^2
 \rho}{\tan^2 \theta + \tanh^2\rho}d\phi_1^2+\frac{1}{\tan^2
 \theta + \tanh^2\rho}d\phi_2^2,\right) \cr
 & B=\frac{k \tanh^2\rho }{\tan^2\theta +
 \tanh^2\rho},~~~~~~g_s^2=\frac{1}{\cos^2\theta ~
 \cosh^2 \rho (\tan^2\theta + \tanh^2 \rho)}~.}}
When $\rho\rightarrow\infty$ we get back the CHS solution. The
full solution corresponds to a ring of fivebranes in the $B$
plane. A simple way to see this is to note that the dilaton
diverges at $\theta=\rho=0$, which is indeed a point in the $B$
plane (see eq. \rel).

\item{(3)} Combining the backgrounds (1) and (2) leads to
$ \left( \frac{SL(2)_{L^2 k} }{U(1)} \times
\frac{SU(2)_k}{U(1)}\right) /\Bbb Z_k$. The asymptotic radius
is $L \sqrt{k}$ and we also have a cigar. This is the supergravity
description of the deformed fivebranes theory perturbed by \tachh.
The spacetime fields take the form
 \eqn\ssl{ds^2=k\left( L^2(d\rho^2 +\tanh ^2 \rho
 d\tilde{\phi}_1^2)+ d\theta^2 +\tan^2 \theta d\tilde{\phi}_2^2
 \right),~~~~g_s=\frac{1}{\cos\theta ~\cosh\rho},}
with the usual   $\Bbb Z_k$ identifications $(\tilde{\phi}_1,
\tilde{\phi}_2) \sim (\tilde{\phi}_1+2\pi/k,
\tilde{\phi}_2+2\pi/k)$. Following the steps above,  we first
use \trans\ to find
 \eqn\slk{ds^2= kd\theta^2 +L^2 k d\rho^2 +L^2 k \tanh^2 \rho
 d\phi_1^2+\frac{1}{k}(L^2 \tanh^2 \rho +\tan^2\theta)d\phi_2^2
 +2L^2 \tanh^2 \rho d\phi_1 d\phi_2.}
Applying  T-duality we get
 \eqn\alal{\eqalign{
 &ds^2=k\left( d\theta^2+L^2 d\rho^2+\frac{L^2 \tan^2\theta
 \tanh^2 \rho}{L^2 \tanh^2 \rho+\tan^2\theta} d\phi_1^2+\frac{1}{L^2
 \tanh^2 \rho+\tan^2\theta}d\phi_2^2\right)~, \cr
 & B=\frac{L^2 \tanh^2 \rho}{L^2 \tanh^2
 \rho+\tan^2\theta},~~~~~g_s^2=\frac{1}{\cos^2\theta \cosh^2\rho (L^2
 \tanh^2 \rho+\tan^2\theta)}~.}}
Now we can write down the supergravity solution that corresponds to the
deformation \tachh\
 \eqn\defo{\eqalign{& ds^2=-dt^2_{new}+d\phi_{\rm new}^2+dx_{||}^2+\cr
 & \qquad
 k\left( d\theta^2+\frac{L^2 \tan^2\theta
 \tanh^2 (\phi_{\rm new}/L\sqrt{k})}{L^2 \tanh^2
(\phi_{\rm new}/L\sqrt{k})+\tan^2\theta} d\phi_1^2+
 \frac{1}{L^2 \tanh^2
 (\phi_{\rm new}/L\sqrt{k})+\tan^2\theta}d\phi_2^2\right)~,\cr
 &
 B=\frac{L^2 k \tanh^2 (\phi_{\rm new}/L\sqrt{k})}{L^2
 \tanh^2(\phi_{\rm new}/L\sqrt{k})+\tan^2\theta}~,\cr
 &
 g_s^2=\frac{1}{\cos^2\theta \cosh^2(\phi_{\rm new}/L\sqrt{k})
 (L^2 \tanh^2 (\phi_{\rm new}/L\sqrt{k})+\tan^2\theta)\exp(2\alpha
 t_{\rm new})}~, \cr }}
where
 \eqn\eqtwo{\alpha^2=\frac{1}{k}(\frac{1}{L^2}-1)}
and $L<1$. A simple way to  find the relation between $t_{\rm
new}, \phi_{\rm new}$ and the original coordinates $t, \phi$  is
to note that asymptotically the background is not deformed. This
gives
 \eqn\roror{ \phi=\frac{1}{L}\phi_{\rm new}+\sqrt{k}\alpha
 t_{\rm new}~, ~~~~t=\frac{1}{L}t_{\rm new}+\sqrt{k}\alpha
 \phi_{\rm new}~,}
in agreement with \defphi .

To verify that this solution indeed describes a ring of $NS5$-branes
which run away to infinity we note that the dilaton diverges when
$\theta=\phi_{\rm new}=0$. In terms of the original coordinates, the
trajectory of the fivebranes  is
\eqn\trajfive{\phi=\left( \frac{1}{L}-L \right) t~.}

\item{(4)} Now we wish to find the supergravity solution that
corresponds to \timedep . For this we have to  proceed in a
slightly different way. At large $\rho$ the metric  takes the form
 \eqn\eqth{ds^2= k d\rho^2 +dx^2 -dt^2 +...}
where $x\sim x+2\pi R$ and $R=L\sqrt{k}$ with $L<1$. Clearly we
can define a new coordinates system obtained by a boost
 \eqn\eqfo{ x_{\rm new}= C x + S t~, ~~~~~t_{\rm new}= C t + S x~,
 ~~~~~~C^2-S^2=1}
so that the metric at infinity takes the form
\eqn\uir{ ds^2= k d\rho^2 +dx_{\rm new}^2
-dt_{\rm new}^2 +\cdots~.}
Now we can deform \uir\ to obtain a cigar like geometry
 \eqn\fqr{ ds^2=k d\rho^2 +\tanh^2\rho\ dx_{\rm new}^2
 -dt_{\rm new}^2 +...}
Since $x$ is periodic  this cannot be done  for any boost
parameter, $C$. To avoid a conical singularity we must impose
 \eqn\cos{C^2=\frac{k}{R^2}=\frac{1}{L^2}~. }
This condition is equivalent by T-duality to \ynewdef. Writing
\fqr\ using $t$ and $\tilde{\phi}_1=x/R$ we get
 \eqn\mea{\eqalign{ds^2=&dx_{||}^2+k(d\rho^2 +d\theta^2 +\tan^2 \theta
 d\tilde{\phi}_2^2)+d\tilde{\phi}_1^2(k \tanh^2 \rho -S^2 R^2)\cr
 &-dt^2(C^2-\tanh ^2 \rho S^2) +2 d\tilde{\phi_1} dt CS R (\tanh^2
 \rho -1)~.\cr}}
Now we can follow the same steps as above. First we change
coordinate from $\tilde{\phi}_i$ to $\phi_i$. This gives (with the
help of \cos)
 \eqn\wrl{\eqalign{ds^2=& dx_{||}^2+k(d\rho^2 +d\theta^2) +
 k(\tanh^2\rho -\epsilon)d\tilde{\phi}_1^2-\frac{1}{L^2}(1-\epsilon
 \tanh^2 \rho)dt^2+ \cr &
 \frac1k (\tanh^2 \rho  +  \tan^2 \theta -\epsilon)d\tilde{\phi}_2^2
 +
2(\tanh^2\rho -\epsilon)d\tilde{\phi}_1 d\tilde{\phi}_2 \cr & -2
\frac{\sqrt{\epsilon k}}{L} \frac{1}{\cosh^2 \rho}dt d
\tilde{\phi}_1-{ 2 \over L} \sqrt{\frac{\epsilon }{k}}
\frac{1}{\cosh^2
 \rho}dt d \tilde{\phi}_2~,\cr }}
where $\epsilon=1- L^2$. Applying  T-duality in the $\phi_2$
direction we get
 \eqn\yuw{\eqalign{
 \tilde{g}_{22}&=\frac{1}{g_{22}}=\frac{k}{\tanh^2 \rho  +  \tan^2
\theta -\epsilon}~,\cr
 \tilde{g}_{11}&=\frac{1}{g_{22}}(g_{11}g_{22}-g_{12}^2)=\frac{k\tan ^2
\theta (\tanh^2 \rho
 -\epsilon)}{\tanh^2 \rho  +  \tan^2 \theta -\epsilon}~,\cr
 \tilde{g}_{tt}&=\frac{1}{g_{22}}(g_{tt}g_{22}-g_{t2}^2)=
 -\frac{\tan^2\theta +\tanh^2\rho (L^4-\epsilon \tan^2 \theta)
 }{L^2
 (\tanh^2 \rho  +  \tan^2 \theta -\epsilon)}~,\cr
 \tilde{g}_{t1}
&=\frac{1}{g_{22}}(g_{t1}g_{22}-g_{t2}g_{12})=\sqrt{\frac{\epsilon k}{L^2}}
 \frac{\tan^2\theta}{\cosh^2 \rho (\tanh^2 \rho  +  \tan^2 \theta
 -\epsilon)}~,\cr
 \tilde{B}_{t2}&=\frac{g_{t2}}{g_{22}}=-\sqrt{\frac{\epsilon k}{L^2}}
 \frac{1}{\cosh^2 \rho(\tanh^2 \rho  +  \tan^2 \theta -\epsilon)}~,\cr
 \tilde{B}_{12}&=\frac{g_{12}}{g_{22}}=k \tilde{B}_{t2}~.\cr}}
This solution describes rotation in the $A$ plane (as opposed to
the previous case where the branes move in the $B$ plane) since
$g_{t2}$ vanishes and $g_{t1}$ does not.

\listrefs
\end